\documentclass[aps,prb,twocolumn,superscriptaddress]{revtex4-2}
\usepackage{amsmath,amssymb,amsfonts}
\usepackage{bm, latexsym}
\usepackage{graphicx}
\usepackage{bbm,times}
\usepackage[normalem]{ulem}
\usepackage[usenames,dvipsnames]{color}
\usepackage[pdftex]{hyperref}
\usepackage{braket}

\begin{document}

\newcommand{\cred}[1]{\textcolor{black}{#1}}
\newcommand{\credrev}[1]{\textcolor{black}{#1}}

%Title of paper
\title{Correlation versus dissipation in a non-Hermitian Anderson impurity model}
\author{Kazuki Yamamoto}
\email{yamamoto@phys.titech.ac.jp}
\affiliation{Department of Physics, Tokyo Institute of Technology, Meguro, Tokyo 152-8551, Japan}
\author{Masaya Nakagawa}
\affiliation{Department of Physics, University of Tokyo, Hongo, Tokyo 113-0033, Japan}
\author{Norio Kawakami}
\affiliation{Fundamental Quantum Science Program, TRIP Headquarters, RIKEN, Wako 351-0198, Japan}
\affiliation{Department of Physics, Ritsumeikan University, Kusatsu, Shiga 525-8577, Japan}
\affiliation{Department of Materials Engineering Science, Osaka University, Toyonaka, Osaka 560-8531, Japan}

\date{\today}

\begin{abstract}
We analyze the competition between strong correlations and dissipation in quantum impurity systems from the Kondo regime to the valence fluctuation regime by developing a slave-boson theory for a non-Hermitian Anderson impurity model with one-body loss. Notably, in the non-Hermitian Kondo regime, strong correlations qualitatively change the nature of dissipation through renormalization effects, where the effective one-body loss is suppressed and emergent many-body dissipation characterized by the complex-valued hybridization is generated. We unveil the mechanism of a dissipative quantum phase transition of the Kondo state on the basis of this renormalization effect, which counterintuitively enhances the lifetime of the impurity against loss. We also find a crossover from the non-Hermitian Kondo regime to the valence fluctuation regime dominated by one-body dissipation. Our results can be tested in a wide variety of setups such as quantum dots coupled to electronic leads and quantum point contacts in ultracold Fermi gases.
\end{abstract}

\maketitle

%%%-----[Introduction]-----
\section{Introduction}
Strong correlations originate from the electron-electron interactions and give rise to exotic phenomena in quantum materials. The Kondo effect, which arises from strong interactions between localized impurity spins and conduction electrons \cite{Kondo64, Hewson97, Coleman15}, has been one of the central problems in condensed matter physics \cite{Schroder00, Aynajian12}, nano science and technology \cite{Gordon98, Cronenwett98, Borzenets20}, and atomic, molecular, and optical (AMO) physics \cite{Gorshkov10, Riegger18}. \credrev{The physics of an impurity system} shows a crossover from the Kondo regime to the valence fluctuation regime depending on the magnitude of charge fluctuations at an impurity site; an impurity fermion is almost localized and forms a Kondo singlet with conduction electrons in the former, while charge fluctuations are significant in the latter. In the Kondo effect, the renormalization effect, which enables an effective description of many-body physics in terms of renormalized one-body parameters, plays a key role. \credrev{As found in a standard textbook \cite{AGD63}, this effect is typically expressed in terms of the renormalization factor, which is given by the frequency-linear term of the self-energy in the Green function and characterizes the width of the Kondo resonance formed around the Fermi energy \cite{Coleman15}.} Owing to the experimental development in quantum simulations with AMO systems \cite{Bloch08, Bloch12, Takahashi20, Konotop16} as well as in the dynamical control of materials with external driving forces \cite{Basov11}, the scope of strong correlation physics has been expanded over various nonequilibrium phenomena in the past couple of decades. Nonequilibrium properties of the Kondo effect have attracted broad interest both theoretically and experimentally, such as in transport through a quantum dot \cite{Meir92, Meir93, Ralph94, Lopez03}, nonlinear conductance in two-channel Kondo problems \cite{Ralph92, Hettler94, Aguado00}, optical absorptions by a quantum quench \cite{Tureci11, Latta11}, and exotic phenomena under time-dependent external fields \cite{Hettler95, Nordlander99, Shahbazyan00, Heyl10, Sbierski13, Nakagawa15}.

Meanwhile, open quantum systems have witnessed a remarkable development in recent years \cite{Muller12, Daley14, Diehl08NP, Yamamoto21, Garcia09, Kraus08, Boite13, Yamamoto20, Yamamoto23B}. In particular, non-Hermitian (NH) physics naturally arises by employing postselections of particular measurement outcomes \cite{Ashida20} and has been intensively investigated thanks to the advancement in dissipation engineering with ultracold atoms \cite{Syassen08, Yan13, Patil15, Schneider17, Tomita17, Spon18, Gerbier20, Honda23, Ott13, Labouvie16, Benary22, Takasu20, Ren22, Liang22, Tsuno24, Jo25}. \cred{So far, NH physics has uncovered unique many-body quantum phenomena \credrev{\cite{Meden23, Nakagawa20, Gopa21, Yang21, Sun23, Yamamoto19, Takemori24, Ghatak18, Hongchao23, Hamazaki19, Hanai19, Nakagawa21, Dora19, Dora20, Moca21, Sticlet22}}, such as unconventional power-law correlations in critical phenomena \cite{Ashida16, Ashida17, Yamamoto22, Yamamoto23}, anomalous reversion of the renormalization group flow in Kondo problems \cite{Nakagawa18, Lourenco18, Han23}, and NH quantum phase transitions in the Kondo effect \cite{Nakagawa18, Hasegawa21, Kattel24, Andrei24, Kulkarni22, NHKondo}.} Among the quantum impurity problems in open quantum systems, lossy quantum dots coupled to fermion reservoirs have been widely explored, and such a local dissipation has been realized in experiments \cite{Esslinger19L, Esslinger19A, Esslinger23}. \cred{Though such a lossy quantum dot has been theoretically well studied in noninteracting problems \credrev{\cite{Jin20, Visuri22, Visuri23, Uchino22, Cazalilla23, Ferreira24,Yoshimura20}}, comprehensive understanding of many-body effects on a quantum dot due to strong correlations and dissipation has not been obtained yet \cite{Dorda14, Dorda15, Schwarz16,Schiro19C, Schiro24, Schiro24arXiv, Stefanini24}.} As the many-body physics of the Kondo effect in closed systems is well understood, here we ask the following question for open systems: how does the renormalization effect due to strong correlations affect the nature of dissipation?

In this paper, we analyze the competition between strong correlations and dissipation in quantum impurity systems by taking the NH Anderson impurity model (NH-AIM) with one-body loss as a prototypical example. We formulate a slave-boson (SB) mean-field theory for open quantum systems by generalizing both the SB field and the renormalized impurity level to complex values. In the NH Kondo regime, we demonstrate that the effective one-body loss is suppressed and converted to emergent many-body loss characterized by the complex-valued hybridization, which highlights that strong correlations qualitatively change the essential nature of dissipation. As a result, the renormalized many-body dissipation induces NH quantum phase transitions with the breakdown of the NH Kondo effect. This result is counterintuitive because the one-body loss usually shortens the lifetime of the impurity, while that in the NH Kondo regime enhances the lifetime through the renormalization effect. Moreover, we study a crossover from the NH Kondo regime to the valence fluctuation regime and show that ramping up the impurity level leads to the suppression of the impurity lifetime caused by one-body dissipation.

The rest of this paper is organized as follows. We introduce the AIM in Sec.~\ref{sec_model} and analyze the noninteracting limit in Sec.~\ref{sec_resonant} to elucidate the effects of one-body loss in an impurity model without local interaction. Then, we formulate the NH SB mean-field theory in Sec.~\ref{sec_NHSB} for the infinite-$U$ NH-AIM. We study the strong correlation effect on the NH-AIM in Sec.~\ref{sec_result}. Finally, we summarize the results in Sec.~\ref{sec_conclusion}

\section{Model}
\label{sec_model}
\begin{figure}[t]
\includegraphics[width=8.5cm]{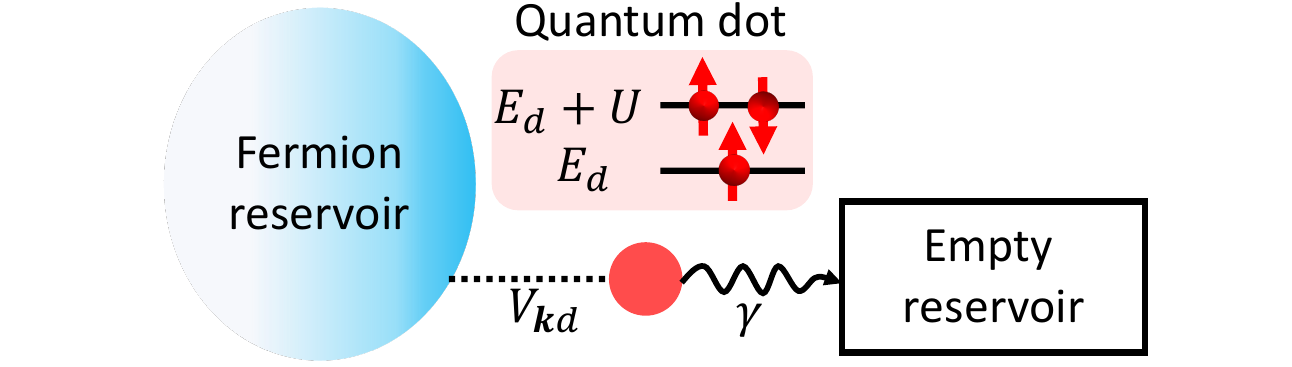}
\caption{\cred{Schematic figure of a quantum dot described by the AIM with one-body loss. Two impurity fermions with spin up and down in the quantum dot with the energy level $E_d$ feel the interaction $U$. The rate of the tunneling to a fermion reservoir is given by $V_{\bm k d}$ and its Hermitian conjugate. Experimentally, in ultracold atoms, the one-body loss from a quantum dot can be controlled by irradiating a tightly focused beam to a quantum point contact \cite{Esslinger23}. In semiconductor quantum dots, the empty reservoir can be realized with an electronic lead with much lower voltage bias compared to that for the fermion reservoir.}}
\label{fig_Experiment}
\end{figure}

We consider the AIM that includes a strong on-site interaction at an impurity site and a coupling to a fermion reservoir. As this model captures essential properties of strongly correlated phenomena, it has been studied for a wide variety of setups in, e.g., localized magnetic moments in metals \cite{Anderson61, Wiegmann80, Kawakami81, Tsvelick83}, quantum dots coupled to electronic leads \cite{Meir91, Beenakker91}, and impurity bound states in mixtures of ultracold atoms \cite{Demler13}. For a single impurity, the AIM is given by
\begin{align}
H=&\sum_{\bm k \sigma}\epsilon_{\bm k} c_{\bm k \sigma}^\dag c_{\bm k \sigma} +\sum_\sigma E_d  n_{d\sigma}\notag\\
&+ U n_{d\uparrow} n_{d\downarrow} + \sum_{\bm k \sigma}\left[V_{\bm k d} c_{\bm k \sigma}^\dag c_{d\sigma} + V_{d \bm k} c_{d\sigma}^\dag c_{\bm k \sigma}\right],
\label{eq_Anderson}
\end{align}
where $c_{d\sigma}$ and $c_{\bm k\sigma}$ denote annihilation operators for fermions at an impurity site and in a fermion reservoir, $n_{d\sigma}=c_{d\sigma}^\dag c_{d\sigma}$ is the particle-number operator at the impurity site, and the hopping rate satisfies $V_{d \bm k}=V_{\bm k d}^*$. We set the chemical potential of the fermion reservoir to zero. In order to study the influence of dissipation, we introduce one-body loss at the impurity site \cite{Wolff20, Froml19, Froml20, Muller21}, which is schematically shown in Fig.~\ref{fig_Experiment}. Within the Markovian approximation, the dynamics of the system density matrix $\rho$ is described by the Lindblad equation \cite{Lindblad76, Gorini76, Breuer02}
\begin{align}
\frac{d\rho}{dt}&=-i[H, \rho] - \frac{\gamma}{2}\sum_\sigma\left(\{L_\sigma^\dag L_\sigma, \rho\}-2L_\sigma\rho L_\sigma^\dag\right)\notag\\
&=-i(H_{\mathrm{eff}}\rho-\rho{H}_{\mathrm{eff}}^\dagger)+\gamma\sum_\sigma{L}_\sigma\rho{L}_\sigma^\dagger,
\label{eq_Lindblad}
\end{align}
where the Lindblad operator $L_\sigma=c_{d\sigma}$ denotes the one-body loss at the impurity site with the rate $\gamma>0$. We here focus on surviving impurity fermions at a short timescale and extract the physics governed by the effective NH Hamiltonian $H_\mathrm{eff}=H-(i\gamma/2)\sum_\sigma L_\sigma^\dag L_\sigma$ \cite{Nakagawa18, Lourenco18, Hasegawa21, Kulkarni22, Han23, Kattel24, Andrei24}. Experimentally, the NH Hamiltonian dynamics is realized by monitoring the empty reservoir and considering the timescale before a fermion is detected. As the Liouvillian spectrum in lossy quantum systems are obtained from the NH Hamiltonian, our theory would be relevant not only to the NH physics but also to the Lindblad dynamics \cite{Nakagawa21}.

\section{Analysis of the non-Hermitian resonant level model}
\label{sec_resonant}
Before analyzing the many-body problem of the NH-AIM, we first study the noninteracting limit ($U\to0$), where the model reduces to the NH resonant level model. The effective Hamiltonian in the Fock space is given by 
\begin{align}
\mathcal H_{0,\mathrm{eff}}=&\sum_{\bm k \sigma}\epsilon_{\bm k} c_{\bm k \sigma}^\dag c_{\bm k \sigma} +\sum_\sigma \tilde E_d  n_{d\sigma} \notag\\
&+ \sum_{\bm k \sigma}\left[V_{\bm k d} c_{\bm k \sigma}^\dag c_{d\sigma} + V_{d \bm k} c_{d\sigma}^\dag c_{\bm k \sigma}\right],
\label{eq_Anderson}
\end{align}
where $\tilde E_d =E_d-i\gamma/2$. We introduce the matrix representation of the single-particle retarded NH Green functions corresponding to Eq.~\eqref{eq_Anderson} as
\begin{align}
G^R(\epsilon)=\frac{1}{\epsilon+i\eta-H_{0,\mathrm{eff}}}.
\label{eq_GreenOperator}
\end{align}
Here, the limit $\eta\to+0$ is implicitly indicated. We find the following relation for the matrix elements of the Green function:
\begin{align}
\sum_\nu(\epsilon+i\eta-H_{0,\mathrm{eff}})_{\mu\nu}G_{\nu\kappa}^R=\delta_{\mu\kappa},\label{eq_Green0_matrix}
\end{align}
where $\delta_{\mu\kappa}$ is the Kronecker delta. Equation \eqref{eq_Green0_matrix} is rewritten as
\begin{gather}
\sum_\nu(\epsilon+i\eta-H_{0,\mathrm{eff}})_{d\nu}G_{\nu d}^R=1\notag\\
\iff (\epsilon+i\eta-\tilde E_d)G_{dd}^{R\sigma}-\sum_{\bm k}V_{d \bm k}G_{\bm k d}^{R\sigma}=1\label{eq_Gdd},
\end{gather}
\begin{gather}
\sum_\nu(\epsilon+i\eta-H_{0,\mathrm{eff}})_{\bm k\nu}G_{\nu d}^R=0\notag\\
\iff (\epsilon+i\eta-\epsilon_{\bm k})G_{\bm k d}^{R\sigma}-V_{\bm k d}G_{dd}^{R\sigma}=0\label{eq_Gkd},
\end{gather}
\begin{gather}
\sum_\nu(\epsilon+i\eta-H_{0,\mathrm{eff}})_{d\nu}G_{\nu \bm k}^R=0\notag\\
\iff (\epsilon+i\eta-\tilde E_d)G_{d\bm k}^{R\sigma}-\sum_{\bm k^\prime}V_{d \bm k^\prime}G_{\bm k^\prime \bm k}^{R\sigma}=0,
\end{gather}
\begin{gather}
\sum_\nu(\epsilon+i\eta-H_{0,\mathrm{eff}})_{\bm k\nu}G_{\nu \bm k^\prime}^R=\delta_{\bm k \bm k^\prime}\notag\\
\iff (\epsilon+i\eta-\epsilon_{\bm k})G_{\bm k \bm k^\prime}^{R\sigma}-V_{\bm k d}G_{d\bm k^\prime}^{R\sigma}=\delta_{\bm k \bm k^\prime},
\end{gather}
where $dd$, $\bm k d$, $d \bm k$, and $\bm k \bm k^\prime$ components of Eq.~\eqref{eq_Green0_matrix} are shown from top to bottom. By substituting Eq.~\eqref{eq_Gkd} into Eq.~\eqref{eq_Gdd}, the Green function $G_{dd}^{R\sigma}(\epsilon)$ of the impurity is calculated as
\begin{align}
G_{dd}^{R\sigma}(\epsilon)=\frac{1}{\epsilon+i\eta-\tilde E_d-\sum_{\bm k}[|V_{d\bm k}|^2/(\epsilon+i\eta-\epsilon_{\bm k})]}.\label{eq_green0i}
\end{align}
We evaluate the lifetime of the impurity from the Green function \eqref{eq_green0i}. Since the real part of the energy shift is not related to the lifetime, which is extracted from the imaginary part of the energy shift, we can rewrite Eq.~\eqref{eq_green0i} as
\begin{align}
G_{dd}^{R\sigma}(\epsilon)=\frac{1}{\epsilon-\tilde E_d+i\Delta}=\frac{1}{\epsilon- E_d+i(\Delta+\frac{\gamma}{2})},
\end{align}
where we have introduced the width of the impurity level in the Hermitian limit as
\begin{align}
\Delta=\pi\sum_{\bm k}|V_{d\bm k}|^2\delta(\epsilon-\epsilon_{\bm k})\sim\pi V^2\rho(\epsilon)\sim\pi V^2\rho_0.
\label{eq_d_width}
\end{align}
Here, $\rho(\epsilon)=\sum_{\bm k}\delta(\epsilon-\epsilon_{\bm k})$ is the density of states of fermions in the reservoir and replaced by the value $\rho_0$ at the Fermi level. The squared hybridization strength $V^2$ is defined by the mean value of $|V_{d\bm k}|^2$ satisfying $\epsilon_{\bm k}=\epsilon$. From the retarded Green function, we obtain the effective density of states as
\begin{align}
\rho_{d\sigma}(\epsilon)=-\frac{1}{\pi}\mathrm{Im}G_{dd}^{R\sigma}(\epsilon)=\frac{1}{\pi}\frac{\Delta+\frac{\gamma}{2}}{(\epsilon-E_d)^2+(\Delta+\frac{\gamma}{2})^2},
\label{eq_DOS_ground}
\end{align}
which is a Lorentzian function. We find in Eq.~\eqref{eq_DOS_ground} that the width of the impurity level $\Delta$ is broadened by dissipation as $\Delta+\frac{\gamma}{2}$, which means that the lifetime of the impurity is reduced because of one-body loss in the NH resonant level model.

\section{Formulation: Non-Hermitian slave-boson mean-field theory}
\label{sec_NHSB}
We now study the interacting case with the use of SB theory generalized to NH systems. Slave-particle theory is a powerful method to analyze strongly correlated phenomena such as the high-temperature superconductivity, spin liquids, and Kondo problems \cite{Lee06, Coleman15, Wingreen94}. We employ the SB theory for the AIM \cite{Coleman84, Coleman87, Newns87} and generalize it to the NH case by letting both the SB field and the Lagrange multiplier be complex. Physically, the Lagrange multiplier enforces a constraint for the total particle number at the impurity site. We note that the SB approach to a $\mathcal{PT}$-symmetric NH Anderson Hamiltonian with real SB fields and real Lagrange parameters has been studied in Ref.~\cite{Kulkarni22}, but our formalism is not restricted to particular dissipation that satisfies the $\mathcal{PT}$ symmetry \cred{\cite{PTAnderson}}.

We focus on the infinite-$U$ limit in the NH-AIM $H_\mathrm{eff}$, where the SB field $b$ is introduced in a similar way to the Hermitian case as $c_{d\sigma}=b^\dag d_\sigma$ with a constraint
\begin{align}
\sum_\sigma d_\sigma^\dag d_\sigma +b^\dag b = 1,
\label{eq_pnconservation}
\end{align}
in order to prohibit the double occupancy of the impurity fermion $d_\sigma$ \cite{Footnote}. To analyze the property of an eigenstate of $H_\mathrm{eff}$ with the smallest real part of energy (see below), we define the partition function in NH systems as $Z=\sum_n e^{-\beta E_n}=\sum_n {}_L\langle E_n|e^{-\beta H_\mathrm{eff}}|E_n\rangle_R$, where $|E_n\rangle_R$ and $|E_n\rangle_L$ are the right and left eigenstates of $H_\mathrm{eff}$ with eigenenergy $E_n$. Here, the eigenstates satisfy the orthonormal relation ${}_L\langle E_m|E_n\rangle_R=\delta_{mn}$. We start from the path-integral representation of the partition function with a constraint as $Z=\int \mathcal D[\bar \psi, \psi, \bar b, b, \tilde \lambda]e^{-S}$, where the action $S$ is given by
\begin{align}
S&=\int_0^\beta d\tau \{\sum_{\bm k\sigma}\bar c_{\bm k \sigma}(\partial_\tau+\epsilon_{\bm k})c_{\bm k \sigma}+\sum_{\sigma}\bar d_{\sigma}(\partial_\tau+\tilde E_d+\tilde \lambda)d_{\sigma}\notag\\
&+\sum_{\bm k \sigma}[V_{\bm k d} \bar c_{\bm k \sigma} \bar b d_\sigma + V_{d\bm k} \bar d_\sigma b c_{\bm k \sigma}]+\bar b (\partial_\tau + \tilde \lambda)b - \tilde \lambda\}.
\label{eq_pathintegral}
\end{align}
Here, $\psi$ denotes the set of fermion fields and $\tilde\lambda$ is the complex-valued Lagrange multiplier. \cred{We consider an energy eigenstate with the smallest real part of the eigenvalue by adiabatically introducing dissipation to the ground state in the Hermitian limit and thus take $\beta\to\infty$ when we discuss physical properties.} We emphasize that the SB field $b$ and the Lagrange multiplier $\tilde \lambda$ should be in general introduced as complex-valued parameters in the NH case. In order to find the saddle point solution, we first integrate out the fermion fields, obtaining the effective action $S_\mathrm{eff}$ as $Z=\int\mathcal D[\bar b, b, \tilde \lambda]e^{-S_\mathrm{eff}}$. Then, the saddle-point condition is given by
\begin{align}
\frac{\delta S_\mathrm{eff}}{\delta \bar b}&=(\partial_\tau+\tilde \lambda) b +\sum_{\bm k \sigma}V_{\bm k d}{}_L\langle c_{\bm k \sigma}^\dag d_\sigma\rangle_R=0,\label{eq_saddle_b2}\\
\frac{\delta S_\mathrm{eff}}{\delta \tilde \lambda}&=\sum_\sigma{}_L\langle d_\sigma^\dag d_\sigma\rangle_R+\bar {b} b-1=0,\label{eq_saddle_lambda2}
\end{align}
where
\begin{gather}
{}_L\langle c_{\bm k \sigma}^\dag d_\sigma\rangle_R=\frac{1}{Z^\prime}\int\mathcal D[\bar \psi, \psi]\bar c_{\bm k \sigma} d_\sigma e^{-S^\prime},\\
{}_L\langle d_\sigma^\dag d_\sigma\rangle_R=\frac{1}{Z^\prime}\int\mathcal D[\bar \psi, \psi]\bar d_\sigma d_\sigma e^{-S^\prime},\\
Z^\prime=\int\mathcal D[\bar \psi, \psi]e^{-S^\prime}.
\end{gather}
Here, $S^\prime$ is obtained by replacing the variables $\bar b$, $b$, and $\tilde \lambda$ in Eq.~\eqref{eq_pathintegral} by their fixed values, and the subscripts $L$ and $R$ stand for the fact that the left and right eigenstates of $H_\mathrm{eff}$ are different from each other \cite{Memo}. The trivial solution $b=0$ always satisfies the self-consistent equations (SCEs) \eqref{eq_saddle_b2} and \eqref{eq_saddle_lambda2}. It is worth noting that the SB fields are written as
\begin{align}
b &=b_0 e^{i\theta}, \label{eq_b}\\
\bar{b}&=b_0 e^{-i\theta},\label{eq_bbar}
\end{align}
where $b_0 \in\mathbb C$ reflecting non-Hermiticity and $\theta$ stands for the U(1) phase \cite{Symmetry, Yamamoto19}. Therefore, $b$ and $\bar b$ are not necessarily complex conjugate to each other in NH physics. \credrev{We also remark that the quantity ${}_L\langle d_\sigma^\dag d_\sigma\rangle_R$ can be a complex number in general though the total number conservation \eqref{eq_pnconservation} always holds. Therefore, the quantity ${}_L\langle d_\sigma^\dag d_\sigma\rangle_R$ should not be regarded as a physical quantity that can be measured. However, it should be emphasized that, if a Hermitian operator $\hat O$ \credrev{commutes with $H_\mathrm{eff}$ and has a simultaneous eigenstate with $H_\mathrm{eff}$}, ${}_L\langle\hat O\rangle_R={}_R\langle\hat O\rangle_R\in \mathbb R$ follows, and such a quantity can be physically observed (see Appendix \ref{sec_phys} for the details).}

\begin{figure}[t]
\includegraphics[width=8.5cm]{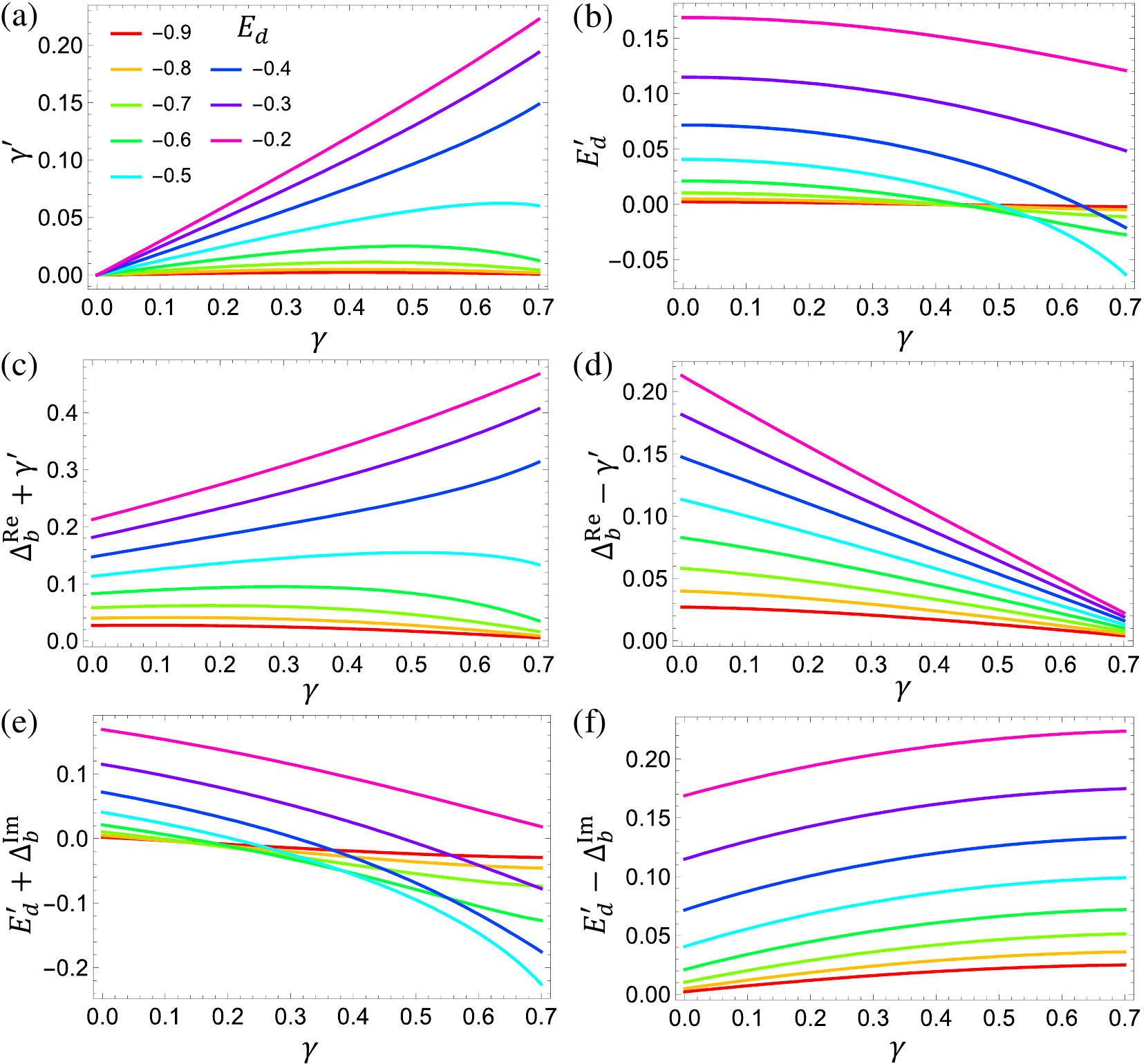}
\caption{\cred{Numerical solutions of the SCE \eqref{eq_self5}. (a) The renormalized one-body loss rate and (b) the renormalized impurity level are almost pinned to zero for $|E_d|\gg\Delta$, highlighting that strong correlations change the qualitative nature of dissipation. (c)[(d)], (e)[(f)] The renormalized resonance width and the renormalized peak position read from $\tilde G_d^{R(A)\sigma}(\omega)$, respectively. The parameters are set to $D=1$, $V=0.5$, and $\Delta=0.39$.}}
\label{fig_KondoResonance}
\end{figure}

In order to solve the SCEs \eqref{eq_saddle_b2} and \eqref{eq_saddle_lambda2}, we assume a static solution for the SB field and calculate the expectation values by formulating NH Green functions as detailed in Appendix \ref{sec_cal}. Then, we find that 
\begin{gather}
{}_L\langle d_\sigma^\dag d_\sigma\rangle_R=\beta^{-1}\sum_{\omega_n} e^{i\omega_n\eta}G_d^\sigma(i\omega_n),\label{eq_Green_d_main}\\
\sum_{\bm k}V_{\bm k d}{}_L\langle c_{\bm k \sigma}^\dag d_\sigma\rangle_R=\beta^{-1}\sum_{\omega_n, \bm k}e^{i\omega_n\eta}V_{\bm k d} G_{d\bm k}^\sigma(i\omega_n),\label{eq_Green_kd_main}
\end{gather}
where $\eta\to+0$ is implied, and the NH Green functions $G_d^\sigma(i\omega_n)$, $G_{d\bm k}^\sigma(i\omega_n)$, and the self-energy $\Sigma_d^\sigma(\omega)$ that is analytically continued to the complex $\omega$ plane are given by
\begin{gather}
G_d^\sigma(i\omega_n)=[i\omega_n-\tilde E_d-\tilde \lambda-\Sigma_d^\sigma(i\omega_n)]^{-1},\label{eq_Green_d_main2}\\
\sum_{\bm k}V_{\bm k d}G_{d \bm k}^\sigma(i\omega_n)=G_d^\sigma(i\omega_n)\Sigma_d^\sigma(i\omega_n)\bar b^{-1},
\end{gather}
and 
\begin{align}
\Sigma_d^\sigma(\omega)=-i \Delta_b \mathrm{sgn}(\mathrm{Im}\omega),\label{eq_SelfEnergy_d_main}
\end{align}
respectively. Here, $\omega_n=(2n+1)\pi/\beta$ with $n\in\mathbb Z$ is the Matsubara frequency for fermions, and $\Delta_b=b_0^2 \Delta$ with $\Delta$ given in Eq.~\eqref{eq_d_width}, where we assume a constant density of states $\rho_0=1/(2D)$ with a cutoff $D$ for fermions in the reservoir. Finally, by carrying out the summation over $\omega_n$ with contour integrations, the SCEs reduce to the following form in the $\beta\to\infty$ limit:
\begin{widetext}
\begin{align}
&\tilde \lambda +\frac{\Delta}{\pi}\log\left[\frac{(E_d^\prime\pm\Delta_b^\mathrm{Im})^2+(\Delta_b^\mathrm{Re}\pm\gamma^\prime)^2}{(D+E_d^\prime\pm\Delta_b^\mathrm{Im})^2+(\Delta_b^\mathrm{Re}\pm\gamma^\prime)^2}\right]
\pm\frac{2i\Delta}{\pi}\left[\tan^{-1}\left(\frac{E_d^\prime\pm\Delta_b^\mathrm{Im}}{\Delta_b^\mathrm{Re}\pm\gamma^\prime}\right)
-\tan^{-1}\left(\frac{D+E_d^\prime\pm\Delta_b^\mathrm{Im}}{\Delta_b^\mathrm{Re}\pm\gamma^\prime}\right)\right]\mp i\Delta_b=\mp i\Delta,
\label{eq_self5}
\end{align}
\end{widetext}
where we have introduced the renormalized impurity level $E_d^\prime\equiv E_d+\mathrm{Re}\tilde\lambda$, (twice) the renormalized one-body loss rate $\gamma^\prime\equiv\gamma/2-\mathrm{Im}\tilde \lambda$, $\Delta_b^\mathrm{Re}\equiv\mathrm{Re}\Delta_b$, and $\Delta_b^\mathrm{Im}\equiv\mathrm{Im}\Delta_b$ \cred{\cite{pmsymbol}}. We point out that Eq.~\eqref{eq_self5} holds for the phase satisfying $\Delta_b^\mathrm{Re}\pm\gamma^\prime>0$, which ensures the analyticity in the half complex plane of the retarded (advanced) Green function given by 
\begin{align}
\tilde G_d^{R(A)\sigma}(\omega)=[\omega-E_d^\prime\mp\Delta_b^\mathrm{Im}\pm i(\Delta_b^\mathrm{Re}\pm\gamma^\prime)]^{-1}.
\label{eq_G_tilde_RA}
\end{align}
As seen from Eq.~\eqref{eq_G_tilde_RA}, the complex-valued hybridization $\Delta_b$ modifies the position of the Kondo peak and the lifetime of the impurity in a nontrivial manner due to non-Hermiticity, which plays a key role in the following. \credrev{As shown below, the vanishment of the resonance width in the Green function correctly signifies the closing of the gap in the real part of the energy spectrum and the jump of the impurity magnetization in the Kondo limit. We note that the energy spectrum and the impurity magnetization are measurable quantities (see Appendix \ref{sec_phys} for the details).}

%%%-----[result]------
\section{Strong correlation effects on the non-Hermitian Anderson impurity model}
\label{sec_result}
\subsection{Non-Hermitian Kondo effect and valence fluctuations}
We numerically solve the SCE \eqref{eq_self5} as shown in Fig.~\ref{fig_KondoResonance}. We find in Figs.~\ref{fig_KondoResonance}(a) and (b) that not only the renormalized impurity level $E_d^\prime$ but also the renormalized one-body loss $\gamma^\prime$ is almost pinned to zero for \cred{$|E_d|\gg\Delta$ ($E_d<0$)}. This result can be seen as a natural but nontrivial generalization of the Kondo physics to NH systems. In the Hermitian Kondo effect, correlation effects renormalize the impurity level to be just above the Fermi level so that its energy scale should be much smaller than the Kondo temperature. On the other hand, in the NH Kondo effect, correlation effects renormalize the complex-valued impurity level so that its amplitude should be much smaller than the typical energy scale of the NH Kondo effect as
\begin{align}
\gamma^\prime, |E_d^\prime|\ll |\Delta_b|.
\end{align}
Thus, even if the one-body loss rate $\gamma$ is increased, correlation effects push back the dissipation to zero and in turn generate the emergent loss characterized by the complex-valued hybridization $\Delta_b$, which indicates that strong correlations qualitatively change the nature of dissipation.

Remarkably, the renormalization of the one-body loss into the many-body dissipation causes NH quantum phase transitions. We define that the phase transition occurs when either the analyticity in the upper or lower-half complex plane of $\tilde G_d^{R(A)\sigma}(\omega)$ breaks down. In Figs.~\ref{fig_KondoResonance}(c) and (d), we find that the renormalized resonance widths $\Delta_b^\mathrm{Re}\pm\gamma^\prime$ obtained from $\tilde G_d^{R(A)\sigma}(\omega)$ are suppressed and vanish with increasing the dissipation $\gamma$ for \cred{$|E_d|\gg\Delta$}. This means that a phase transition from the NH Kondo state to an unscreened impurity spin state emerges. Strictly speaking, the resonance widths obtained from the numerical calculation seem to jump at the dissipation strength just above $\gamma\sim0.7$. However, analytical calculation given around Eq.~\eqref{eq_phasetransition} corroborates the existence of the phase transition characterized by the vanishment of $\Delta_b^\mathrm{Re}\pm\gamma^\prime$ in the Kondo limit. We emphasize that such a transition is counterintuitive because one-body loss in the noninteracting case only suppresses the lifetime of the impurity (see Sec.~\ref{sec_resonant}), while the suppression of the resonance widths rather indicates the enhancement of the lifetime. This result also unveils the origin of the phase transition in the NH Kondo effect \cite{Nakagawa18} from the perspective of the NH-AIM.

For the behavior for the intermediate depth of $E_d$ shown in Fig.~\ref{fig_KondoResonance}(c), the resonance width $\Delta_b^\mathrm{Re}+\gamma^\prime$ is first enhanced with increasing the loss rate and then suppressed. Such a transient loss-induced enhancement of the resonance width is one of the interesting results of the NH SB theory that cannot be seen in the previously studied NH Kondo problems \cite{Nakagawa18, Lourenco18, Hasegawa21, Kulkarni22, Han23, Kattel24, Andrei24}.

Moreover, we find distinct behavior as the impurity level becomes closer to the Fermi level. In Figs.~\ref{fig_KondoResonance}(a) and (b), the deviation of $\gamma^\prime$ and $E_d^\prime$ from zero indicates that the effective one-body loss governs the physics in the valence-fluctuation regime because the many-body effect is too weak to form a Kondo peak. Accordingly, in Fig.~\ref{fig_KondoResonance}(c), we indeed see that ramping up of $E_d$ leads to the enhancement of $\Delta_b^\mathrm{Re}+\gamma^\prime$. These results demonstrate that the NH Kondo regime crossovers to the valence fluctuation regime dominated by one-body dissipation. On the other hand, $\Delta_b^\mathrm{Re}-\gamma^\prime$ in Fig.~\ref{fig_KondoResonance}(d) decreases as a function of $\gamma$ because the dissipation acts as loss of holes, or gain of particles, in the advanced Green function $\tilde G_d^{A\sigma}(\omega)$.

The effect of non-Hermiticity can be also seen in the position of the Kondo peak obtained from $\tilde G_d^{R\sigma}(\omega)$. In Fig.~\ref{fig_KondoResonance}(e), we see for \cred{$|E_d|\gg\Delta$} that the renormalized peak position $E_d^\prime+\Delta_b^\mathrm{Im}$ read from $\tilde G_d^{R\sigma}(\omega)$ is just above the Fermi level in the Hermitian limit, but gradually decreases with changing the sign as the dissipation strength is increased. This behavior is caused by the decrease of $\Delta_b^\mathrm{Im}$, which does not vanish at the transition point. We also find that such a decrease of the peak position becomes significant as we ramp up the impurity level $E_d$. Correspondingly, in Fig.~\ref{fig_KondoResonance}(f), the renormalized peak position $E_d^\prime-\Delta_b^\mathrm{Im}$ for $\tilde G_d^{A\sigma}(\omega)$ is always above the Fermi level and increases further as we introduce dissipation. Since the retarded and advanced Green functions are complex conjugate to each other in the $\gamma\to0$ limit as seen from Eq.~\eqref{eq_G_tilde_RA}, the distinct behavior between them are the unique property in NH physics.

\begin{figure}[t]
\includegraphics[width=8.5cm]{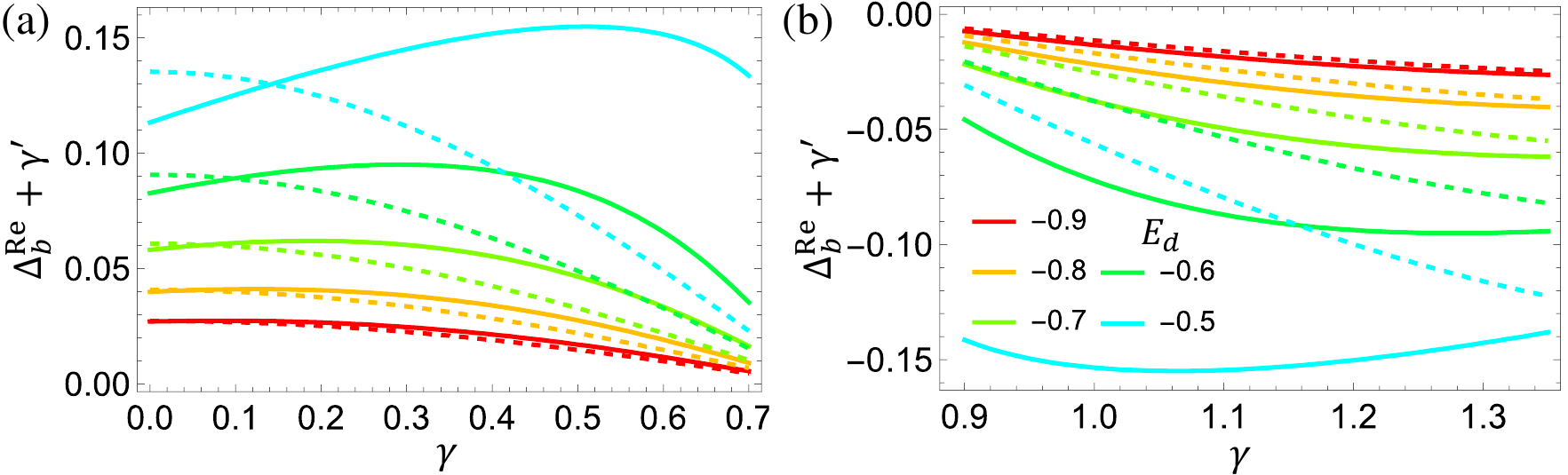}
\caption{\cred{Comparison between $T_K^\mathrm{NH}$ (dashed curves) and the resonance width $\Delta_b^\mathrm{Re}+\gamma^\prime$ (solid curves) obtained from the numerical calculation of Eq.~\eqref{eq_self5} for the NH Kondo state in (a) and for the metastable solution after the phase transition characterized by the \credrev{negative value of the effective density of states} in (b). Analytical results of $T_K^\mathrm{NH}$ capture the behavior of the resonance width for \cred{$|E_d|\gg\Delta$}. The parameters are set to the same values as those in Fig.~\ref{fig_KondoResonance}.}}
\label{fig_KondoTemperature}
\end{figure}

\subsection{Non-Hermitian Kondo scale}
In order to characterize the NH Kondo effect, we generalize the Kondo temperature to a complex energy scale, which we call the NH Kondo scale $\tilde T_K^\mathrm{NH}$. From Eq.~\eqref{eq_self5}, the NH Kondo scale is obtained as an energy scale $\Delta_b$ in the Kondo limit (see Appendix \ref{sec_Kondoscale} for the detailed calculation), given by
\begin{align}
\tilde T_K^\mathrm{NH}=D\exp[\pi\tilde E_d/(2\Delta)],
\label{eq_TKNHtilde_main}
\end{align}
whose real part gives
\begin{align}
T_K^\mathrm{NH}=D\cos[\pi\gamma/(4\Delta)]\exp[\pi E_d/(2\Delta)].\label{eq_NHKondoTemp}
\end{align}
We note that Eqs.~\eqref{eq_TKNHtilde_main} and \eqref{eq_NHKondoTemp} hold in the NH Kondo regime for \cred{$|E_d|\gg\Delta$}. Then, the dissipation-induced quantum phase transition occurs at
\begin{align}
T_K^\mathrm{NH}=0,\label{eq_phasetransition}
\end{align}
or $\gamma=2\Delta$, where $|\mathrm{Im}\tilde E_d|^{-1}$ becomes of the order of the lifetime of the impurity in the noninteracting case. Figure \ref{fig_KondoTemperature}(a) shows the comparison between $T_K^\mathrm{NH}$ given in Eq.~\eqref{eq_NHKondoTemp} and the numerical results of the resonance width $\Delta_b^\mathrm{Re}+\gamma^\prime$. In the NH Kondo regime, where $\gamma^\prime\ll\Delta_b^\mathrm{Re}$ owing to the renormalization effect, we find that the behavior of the resonance width agrees with that of $T_K^\mathrm{NH}$. On the other hand, $\Delta_b^\mathrm{Re}+\gamma^\prime$ deviates from $T_K^\mathrm{NH}$ as the impurity level is raised. Interestingly, we find that Eq.~\eqref{eq_phasetransition} is rewritten as
\begin{align}
[\mathrm{Re}(\tilde E_d^{-1})]^2+[\mathrm{Im}(\tilde E_d^{-1})-(2\Delta)^{-1}]^2=[(2\Delta)^{-1}]^2,\label{eq_circleNHKondo2}
\end{align}
which indicates that the trajectory of the transition point forms a circle in the complex $\tilde E_d^{-1}$ plane. \cred{By replacing the parameter $2\Delta/(\pi\tilde E_d)$ with the complex Kondo coupling, Eq.~\eqref{eq_circleNHKondo2} reduces to the exact result for the NH Kondo model \cite{Nakagawa18}, which vindicates our SB mean-field treatment. Importantly, the exact solution shows that the impurity magnetization, which is a measurable quantity, jumps from $0$ to $1/2$ at the transition. Thus, our theory correctly captures the phase transition that reflects the behavior of physical observables (see Appendix \ref{sec_phys}).} \credrev{We note that the NH phase transition in the valence-fluctuation regime cannot be captured by the NH Kondo model.}

Importantly, in Fig.~\ref{fig_KondoTemperature}(b) for strong dissipation after the phase transition, we find a nontrivial solution of the SCEs \eqref{eq_saddle_b2} and \eqref{eq_saddle_lambda2} characterized by the negative resonance width $\Delta_b^\mathrm{Re}\pm\gamma^\prime<0$. The solution for $\Delta_b^\mathrm{Re}\pm\gamma^\prime<0$ gives a \credrev{negative value of the effective density of states} in the sense of the inverted imaginary part of $\tilde G_d^{R\sigma}(\omega)$ \cite{Kulkarni22}. \credrev{We note that, in nonequilibrium physics, the inversion of the sign of the density of states can occur. For example, it can be caused by the population inversion with respect to the energy eigenstate \cite{Schiro19}.} The solution with the negative resonance width is metastable and the ground state in the sense of the real part of the energy is the localized free spin state characterized by $b_0=0$ (see the next subsection). We also find that the nontrivial solution of the SCEs \eqref{eq_saddle_b2} and \eqref{eq_saddle_lambda2} is allowed only when $\Delta_b^\mathrm{Re}\pm\gamma^\prime>0$ or $\Delta_b^\mathrm{Re}\pm\gamma^\prime<0$ is satisfied. Notably, after the phase transition for $\Delta_b^\mathrm{Re}\pm\gamma^\prime<0$, $\tilde G_d^{R(A)\sigma}(\omega)$ which is defined from analytical continuation of the Matsubara Green function can no longer be regarded as the Fourier transform of the ordinary retarded (advanced) Green function that preserves the causality \cite{Kulkarni22}. In this regime, the SCE \eqref{eq_self5} should be modified to incorporate the finite contribution from poles of $\tilde G_d^{R(A)\sigma}(\omega)$ in contour integrations to calculate Eqs.~\eqref{eq_Green_d_main} and \eqref{eq_Green_kd_main} (see Appendix \ref{sec_pole} for the detailed form). However, Eq.~\eqref{eq_NHKondoTemp} is still valid for $\Delta_b^\mathrm{Re}\pm\gamma^\prime<0$ because the modified SCE for the nontrivial solution reduces to Eq.~\eqref{eq_self5} in the large $D$ limit. This means that such a modification by the pole, which is different from the contribution originating from the integration along the branch cut that typically emerges in impurity physics \cite{Coleman87}, does not affect the low-energy physics of the system. In Fig.~\ref{fig_KondoTemperature}(b), we indeed see that the numerical behavior of the negative resonance width is accurately described by Eq.~\eqref{eq_NHKondoTemp} for \cred{$|E_d|\gg\Delta$}. Importantly, from Eq.~\eqref{eq_saddle_lambda2}, the negative sign of $\Delta_b^\mathrm{Re}$ permits the real part of $\sum_\sigma{}_L\langle d_\sigma^\dag d_\sigma\rangle_R$ to be larger than $1$, which is enabled by non-Hermiticity \cite{Yoshimura20}. Though the region between Figs.~\ref{fig_KondoTemperature}(a) and (b) is not depicted due to the numerical limitation, the resonance width $\Delta_b^\mathrm{Re}+\gamma^\prime$ in the Kondo limit for \cred{$|E_d|\gg\Delta$} is continuously connected as indicated in Eq.~\eqref{eq_NHKondoTemp}, while that in the other region seems to jump at $\gamma=2\Delta$ from positive to negative.

\subsection{Analysis of the stability of the solution}
\label{sec_metastable}
The SCEs \eqref{eq_saddle_b2} and \eqref{eq_saddle_lambda2} have both a nontrivial solution $b_0\neq0$ and a trivial solution $b_0=0$. This means that even if we find a nontrivial solution, it is not guaranteed that it would be a ground state in the sense of the real part of the energy compared to the trivial solution. Here, we investigate the stability of the solution of the SCEs by comparing the real part of the complex-valued energy corresponding to $b_0\neq0$ with that corresponding to $b_0=0$. By taking the expectation value of the effective mean-field Hamiltonian given by
\begin{align}
H_\mathrm{eff}^\mathrm{MF}(\tilde \lambda)=&\sum_{\bm k \sigma} \epsilon_{\bm k} c_{\bm k \sigma}^\dag c_{\bm k \sigma} + \sum_\sigma (\tilde E_d +\tilde \lambda) d_\sigma^\dag d_\sigma \notag\\
&+\sum_{\bm k \sigma}[V_{\bm k d} \bar {b} c_{\bm k \sigma}^\dag d_\sigma + V_{d\bm k}b d_\sigma^\dag c_{\bm k \sigma}]+\tilde \lambda(b_0^2 -1),
\label{eq_HMF}
\end{align}
we find that the energy of the system is evaluated as
\begin{align}
E_\mathrm{eff}^\mathrm{MF}=&\sum_{\bm k\sigma}\epsilon_{\bm k}{}_L\langle c_{\bm k \sigma}^\dag c_{\bm k \sigma}\rangle_R+\sum_\sigma \tilde E_d {}_L\langle d_\sigma^\dag d_\sigma\rangle_R\notag\\
&+\bar{b}\sum_{\bm k \sigma}V_{\bm k d}{}_L\langle c_{\bm k \sigma}^\dag d_\sigma\rangle_R+b\sum_{\bm k \sigma}V_{d\bm k}{}_L\langle d_\sigma^\dag c_{\bm k \sigma}\rangle_R.
\label{eq_energy}
\end{align}
We note that, if the saddle-point condition given in Eqs.~\eqref{eq_saddle_b2} and \eqref{eq_saddle_lambda2} is satisfied, the third term and the forth term in Eq.~\eqref{eq_energy} give the same energy contribution both before and after the phase transition. We evaluate the energy difference between the nontrivial solution and the trivial solution by taking the leading-order contribution of Eq.~\eqref{eq_energy}. By using Eqs.~\eqref{eq_Green_d_cutoff}, \eqref{eq_Green_kd_cutoff}, \eqref{eq_Green_d2_pole}, and \eqref{eq_Green_kd_3_pole} in Appendix \ref{sec_cal} and \ref{sec_pole}, we find
\begin{align}
\mathrm{Re}E_\mathrm{eff}^\mathrm{MF}|_{b_0\neq0}
-\mathrm{Re}E_\mathrm{eff}^\mathrm{MF}|_{b_0=0}
\sim-\frac{2\Delta_b^\mathrm{Re}}{\pi}\log D,
\label{eq_condenergy}
\end{align}
where we have used the fact that the energy scale of the cutoff $D$ is much larger than the other energy scales. The second term in the right-hand side of Eq.~\eqref{eq_energy} does not have the $\log D$ contribution, and Eq.~\eqref{eq_condenergy} comes from the third and the forth terms in the right-hand side of Eq.~\eqref{eq_energy}. Equation \eqref{eq_condenergy} indicates that the nontrivial solution is energetically stable for $\Delta_b^\mathrm{Re}>0$ and is metastable for $\Delta_b^\mathrm{Re}<0$. From Figs.~\ref{fig_KondoTemperature}(a) and (b), we see that $\Delta_b^\mathrm{Re}>0$ should hold before the phase transition and $\Delta_b^\mathrm{Re}<0$ should hold after the phase transition in the Kondo limit. This means that the NH Kondo state for $\gamma<2\Delta$ is stable, but the solution after the phase transition characterized by the \credrev{negative value of the effective density of states} becomes metastable. We also find that further numerical results shown in Appendix \ref{sec_cal} and \ref{sec_pole} [Fig.~\ref{fig_KondoResonance2}(a) and Fig.~\ref{fig_KondoResonanceWithPole}(g)] suggest that all the states ranging from the NH Kondo regime to the valence fluctuation regime are stable before the phase transition for small dissipation and becomes metastable after the phase transition for large dissipation. Then, if we experimentally realize the ground state, the dissipation-induced phase transition from the NH Kondo phase to an unscreened phase can be observed. We also emphasize that the \credrev{negative value of the effective density of states} characterized by the negative resonance width is prohibited in the Hermitian Kondo problem, and thus the existence of such a metastable solution after the dissipation-induced phase transition is a unique property of the NH Kondo problem.

%%%-----[Conclusion]-----
\section{Conclusions and discussions}
\label{sec_conclusion}
We have investigated how strong correlations and dissipation compete in NH impurity physics by formulating the NH SB theory in terms of generalized complex parameters. We have demonstrated in the NH Kondo regime that strong correlations renormalize the one-body loss to zero with invoking the emergent many-body dissipation that induces a quantum phase transition, which indicates the qualitative change of the nature of dissipation. Our results open a new avenue for dissipation engineering of strongly correlated phenomena by using experimental techniques for realizing the AIM in semiconductor quantum dots. The NH-AIM can also be realized with ultracold atoms \cite{Demler13, Nishida16}, e.g., by introducing one-body loss to a quantum point contact \cite{Ono21} and postselecting null measurement outcomes with the use of quantum-gas microscopy \cite{Ott16Rev}. Since lossy fermion systems are also relevant to some solid-state systems \cite{Hanai24}, we believe that our work stimulates further study of dissipative many-body phenomena not only in AMO physics but also in solid-state physics. As anomalous reversion of the renormalization group flow is reported in the NH Kondo model for a specific region of the complex Kondo coupling \cite{Nakagawa18}, it merits further study to explore the relation between the metastable solution obtained in our study and the renormalization group flow.

\begin{acknowledgments}
We are grateful to Yoshiro Takahashi and Koki Ono for the discussion on the experimental setup for impurity models in ultracold atoms. We thank Thierry Giamarchi for fruitful discussions on a dissipative quantum dot. K.Y. thanks Soma Takemori for the detailed discussion on contour integrations for NH systems. This work was supported by KAKENHI Grants No.\ JP20K14383, No.\ JP23K19031, and No.\ JP24K16989. K.Y. was also supported by Yamaguchi Educational and Scholarship Foundation, Toyota RIKEN Scholar Program, Murata Science and Education Foundation, Public Promoting Association Kura Foundation, Hirose Foundation, and the Precise Measurement Technology Promotion Foundation. N.K. was supported by the RIKEN TRIP initiative.
\end{acknowledgments}

%%%%%%%[Supplemental Materials]%%%%%%%%%%

%\clearpage

%\renewcommand{\thesection}{S\arabic{section}}
%\renewcommand{\theequation}{S\arabic{equation}}
%\setcounter{equation}{0}
%\renewcommand{\thefigure}{S\arabic{figure}}
%\setcounter{figure}{0}

%\onecolumngrid

\appendix

\section{Relation to physical quantities and exact results}
\label{sec_phys}
\cred{Here, we explain how our result is related to physical quantities in NH physics. Our analysis of expectation values calculated from left and right eigenstates (which we call LR quantities) is related to the measurable quantity with real values calculated from expectation values with right eigenstates (which we call RR quantities). This is because (complex) LR quantities such as Green functions and (real) RR quantities such as the real and imaginary parts of the energy spectrum and the impurity magnetization both exhibit a singularity at the same critical point. This fact is further supported by the general statement that LR and RR quantities are equal for a Hermitian operator $\hat O$ that commutes with the effective Hamiltonian $H_\mathrm{eff}$.}

\cred{In NH physics, two types of expectation values can be defined according to whether the left or right eigenstate is assigned to the bra vector corresponding to the right eigenstate for the ket vector. For an operator $\hat O$, the former type is defined by ${}_L\langle \hat O \rangle_R\equiv{}_L\langle \psi|\hat O|\psi\rangle_R$, and the latter type is defined by ${}_R\langle \hat O \rangle_R\equiv{}_R\langle \psi|\hat O|\psi\rangle_R$, where the subscripts $L$ and $R$ denote the left and right eigenstate of the effective Hamiltonian $H_\mathrm{eff}=H-\frac{i}{2}\gamma\sum_\sigma L_\sigma^\dag L_\sigma$ as
\begin{gather}
H_\mathrm{eff}|\psi\rangle_R=E|\psi\rangle_R,\\
H_\mathrm{eff}^\dag|\psi\rangle_L=E^*|\psi\rangle_L.
\end{gather}
Here, we have assumed that $|\psi\rangle_L$ and $|\psi\rangle_R$ are normalized as ${}_R\langle\psi|\psi\rangle_R={}_L\langle\psi|\psi\rangle_R=1$ for simplicity. The LR quantity ${}_L\langle\hat O\rangle_R$ takes a complex value and can be calculated with the path-integral formulation dealt with in our study. On the other hand, the RR quantity ${}_R\langle\hat O\rangle_R$ takes a real value and thus is physically measurable.}

\cred{In our study, we evaluate the Green function in order to find a quantum phase transition. The singularity of the Green function at the quantum phase transition point reflects the singularity of the eigenstates $|\psi\rangle_L$ and $|\psi\rangle_R$. This leads to a singular behavior of the energy spectrum, the real and imaginary parts of which are described by measurable (RR) quantities as
\begin{align}
\mathrm{Re} E&=\mathrm{Re}[{}_L\langle\psi|H_\mathrm{eff}|\psi\rangle_R]\notag\\
&=\mathrm{Re}[{}_R\langle\psi|H_\mathrm{eff}|\psi\rangle_R]\notag\\
&={}_R\langle\psi|H|\psi\rangle_R,
\end{align}
and
\begin{align}
\mathrm{Im} E&=\mathrm{Im}[{}_L\langle\psi|H_\mathrm{eff}|\psi\rangle_R]\notag\\
&=\mathrm{Im}[{}_R\langle\psi|H_\mathrm{eff}|\psi\rangle_R]\notag\\
&=-\frac{1}{2}{}_R\langle\psi|\sum_\sigma L_\sigma^\dag L_\sigma|\psi\rangle_R,
\end{align}
which are regarded as the effective energy and the decay rate, respectively. We indeed see that the real part of the energy difference between the NH Kondo state and the unscreened state vanishes at the phase transition point as described in Sec.~\ref{sec_result}, and the NH Kondo state becomes metastable at this point. Thus, the Green function (LR quantity) in our study correctly captures the singularity of the real part of the energy spectrum (RR quantity). We also mention that, in the NH Anderson model considered in our study, the effective Hamiltonian has a symmetry $H_\mathrm{eff}^\dag=H_\mathrm{eff}^*$ and the left eigenstate is obtained by the complex conjugate of the right eigenstate as $|\psi\rangle_L=|\psi\rangle_R^*$. Thus, the behavior of the left eigenstate is closely related to that of the right eigenstate.}

\cred{Furthermore, $|\psi\rangle_L$ and $|\psi\rangle_R$ can be chosen as simultaneous eigenstates of $H_\mathrm{eff}$ and a Hermitian operator $\hat O$ that commutes with $H_\mathrm{eff}$, and thus LR and RR quantities match for such $\hat O$. In fact, if a Hermitian operator $\hat O$ commutes with $H_\mathrm{eff}$, we obtain
\begin{gather}
[H_\mathrm{eff}, \hat O]=0,\label{eq_conS}\\
[H_\mathrm{eff}^\dag, \hat O]=0.
\end{gather}
In the following, we assume that the eigenstates of $H_\mathrm{eff}$ are not degenerate (the eigenstates of $\hat O$ can be degenerate). From Eq.~\eqref{eq_conS}, we obtain
\begin{gather}
H_\mathrm{eff}|\psi\rangle_R=E|\psi\rangle_R\notag\\
\Rightarrow 
\hat OH_\mathrm{eff}|\psi\rangle_R=E\hat O|\psi\rangle_R=H_\mathrm{eff}\hat O|\psi\rangle_R.
\end{gather}
Then, we see that $\hat O|\psi\rangle_R$ is also an eigenstate of $H_\mathrm{eff}$. Because the eigenstate of $H_\mathrm{eff}$ has no degeneracy, we obtain
\begin{align}
\hat O|\psi\rangle_R=M|\psi\rangle_R,
\end{align}
where $M$ is real because $\hat O$ is a Hermitian operator. Thus, $|\psi\rangle_R$ is a simultaneous eigenstate of $H_\mathrm{eff}$ and $\hat O$. Similarly, for the left eigenstate $|\psi\rangle_L$ of the effective Hamiltonian corresponding to $|\psi\rangle_R$, we have
\begin{gather}
H_\mathrm{eff}^\dag |\psi\rangle_L=E^*|\psi\rangle_L\notag\\
\Rightarrow 
\hat OH_\mathrm{eff}^\dag|\psi\rangle_L=E^*\hat O|\psi\rangle_L=H_\mathrm{eff}^\dag \hat O|\psi\rangle_L,
\end{gather}
which leads to
\begin{align}
\hat O|\psi\rangle_L=M|\psi\rangle_L,
\end{align}
where we have used that $|\psi\rangle_L$ is a left eigenstate of $\hat O$ corresponding to $|\psi\rangle_R$. Therefore, we arrive at 
\begin{align}
{}_R\langle\psi|\hat O|\psi\rangle_R={}_L\langle\psi|\hat O|\psi\rangle_R=M,
\end{align}
which is a measurable quantity that reflects the singularity of the energy spectrum. In summary, we have seen that the Green function (LR quantity) can characterize a quantum phase transition, the singularity of which at the transition point reflects a singular behavior of the real and imaginary parts of the energy spectrum (RR quantity), and accordingly the physical observable $\hat O$ (RR quantity). In our study, $\hat O$ corresponds to $\sum_{\bm k} c_{\bm k \sigma}^\dag c_{\bm k \sigma} + d_\sigma^\dag d_\sigma$, which is the total particle number of fermions with spin $\sigma$. In the Kondo limit, the phase transition leads to a jump of the impurity magnetization, which is in fact a measurable quantity. We explain the details in the following.}

\cred{Our result on the quantum phase transition from the NH Kondo phase to an unscreened phase induced by one-body loss agrees with the exact solution of the NH Kondo model obtained in Ref.~\cite{Nakagawa18}. The quantum phase transition from the NH Kondo phase to an unscreened phase occurs at the critical dissipation rate where
\begin{align}
[\mathrm{Re}(\tilde E_d^{-1})]^2+[\mathrm{Im}(\tilde E_d^{-1})-(2\Delta)^{-1}]^2=[(2\Delta)^{-1}]^2 \label{eq_circleNHKondo2S}
\end{align}
[equivalent to Eq.~\eqref{eq_circleNHKondo2}] is satisfied. By replacing the parameter $2\Delta/(\pi \tilde E_d)$ with the complex Kondo coupling, Eq.~\eqref{eq_circleNHKondo2S} reduces to the quantum phase transition point obtained from the exact Bethe ansatz solution of the NH Kondo model.}

\cred{We here summarize the Bethe ansatz analysis studied in Ref.~\cite{Nakagawa18}. When the impurity level is sufficiently deep and satisfies $|E_d|\gg\Delta$, the impurity fermion is localized and the model is described by the NH Kondo Hamiltonian with the complex Kondo coupling as
\begin{align}
H_\mathrm{eff}=&\sum_{\bm k, \sigma}\epsilon_{\bm k}c_{\bm k \sigma}^\dag c_{\bm k \sigma} \notag\\
&+\frac{1}{N_s}\sum_{\bm k, \bm k^\prime, \sigma, \sigma^\prime} c_{\bm k \sigma}^\dag c_{\bm k^\prime \sigma^\prime}(v \delta_{\sigma\sigma^\prime}-J\bm \sigma_{\sigma\sigma^\prime}\cdot \bm S_{\mathrm{imp}}),
\end{align}
where $N_s$ is the number of sites, $\bm \sigma$ is the three-component Pauli matrix vector, $\bm S_\mathrm{imp}$ is the impurity spin operator, $v$ is the complex-valued potential, and $J$ is the complex-valued Kondo coupling. This model is exactly solvable by focusing on the ground state in the sense of the real part of the energy. According to the Bethe ansatz results, the transition point is given by 
\begin{align}
\mathrm{Im}\left(\frac{1}{g}\right)=\frac{1}{2},
\label{eq_magjump}
\end{align}
where $g$ is defined by $g =-\tan(\pi\rho_0 J)$ and $\rho_0$ is the density of states at the Fermi energy. By rewriting Eq.~\eqref{eq_magjump}, we obtain
\begin{align}
\sinh(2\pi\rho_0 J_i)-\frac{1}{2} \cosh(2\pi\rho_0 J_i)=-\frac{1}{2} \cos(2\pi\rho_0 J_r),
\end{align}
where $J_r$ and $J_i$ denote the real and imaginary parts of the Kondo coupling $J$. In the limit $|\rho_0 J_r |\ll1$ and $|\rho_0 J_i |\ll1$, we obtain
\begin{align}
\left(\rho_0 J_r \right)^2+\left(\rho_0 J_i-\frac{1}{\pi}\right)^2=\frac{1}{\pi^2},
\end{align}
which is nothing but Eq.~\eqref{eq_circleNHKondo2S} by replacing $\rho_0 J$ with $2\Delta/(\pi\tilde E_d)$. Importantly, according to the Bethe ansatz study, \credrev{the magnetization operator commutes with $H_\mathrm{eff}$} and the eigenstate of the effective Hamiltonian is a simultaneous eigenstate of the total particle number $n_\sigma^\mathrm{total}$ of spin-$\sigma$ fermions. Here, the impurity magnetization \credrev{in the NH Kondo model} is defined as the difference of the magnetization $m_z^\mathrm{total}$  between the case with an impurity fermion and that without an impurity fermion, where $m_z^\mathrm{total}$ is given by
\begin{align}
m_z^\mathrm{total}=\frac{1}{2}(n_\uparrow^\mathrm{total}-n_\downarrow^\mathrm{total}).
\end{align}
Then, both LR and RR quantities agree for the magnetization, and it is found in Ref.~\cite{Nakagawa18} that the impurity magnetization jumps from $0$ to $1/2$ when we pass through the phase transition point from the NH Kondo phase to an unscreened phase. Therefore, our theory based on the Green function with the use of path integrals correctly captures the phase transition of the physical observable that should be a real quantity.}

\section{Detailed derivation of the self-consistent equation}
\label{sec_cal}
In order to solve the SCEs \eqref{eq_saddle_b2} and \eqref{eq_saddle_lambda2}, we have to calculate the path integral for fixed $b$, $\bar b$, and $\tilde \lambda$ in the partition function
\begin{align}
Z^\prime=\int\mathcal D[\bar \psi, \psi]e^{-S^\prime},
\end{align}
where
\begin{align}
S^\prime=&\int_0^\beta d\tau \Bigg\{\sum_{\bm k\sigma}\bar c_{\bm k \sigma}(\tau)(\partial_\tau+\epsilon_{\bm k})c_{\bm k \sigma}(\tau)\notag\\
&+\sum_{\sigma}\bar d_{\sigma}(\tau)(\partial_\tau+\tilde E_d+\tilde \lambda)d_{\sigma}(\tau)\notag\\
&+\sum_{\bm k \sigma}[V_{\bm k d} \bar b \bar c_{\bm k \sigma}(\tau) d_\sigma(\tau) + V_{d\bm k}  b \bar d_\sigma(\tau) c_{\bm k \sigma}(\tau)]\notag\\
&+\tilde \lambda(b_0^2 -1)\Bigg\},
\label{eq_SMF}
\end{align}
$\psi$ denotes the set of fermion fields, $b_0$ is defined in Eqs.~\eqref{eq_b} and \eqref{eq_bbar},  and we have ignored the imaginary-time dependence of $b$ and $\bar b$ since we are interested in the static solution of the SB fields. We find that the quantities ${}_L\langle d_\sigma^\dag d_\sigma\rangle_R$ and ${}_L\langle c_{\bm k \sigma}^\dag d_\sigma\rangle_R$ are described by the equal-time Green functions as
\begin{gather}
{}_L\langle d_\sigma^\dag d_\sigma\rangle_R=G_d^\sigma(\tau\to-0)=\frac{1}{\beta}\sum_{\omega_n} e^{i\omega_n\eta}G_d^\sigma(i\omega_n),\label{eq_Green_d}\\
{}_L\langle c_{\bm k \sigma}^\dag d_\sigma\rangle_R=G_{d\bm k}^\sigma(\tau\to-0)=\frac{1}{\beta}\sum_{\omega_n} e^{i\omega_n\eta}G_{d\bm k}^\sigma(i\omega_n),\label{eq_Green_kd}
\end{gather}
where the infinitesimal $\eta\to+0$ is introduced as a convergence factor, and we have defined the Green functions in NH systems as
\begin{gather}
G_d^\sigma(\tau, \tau^\prime)=-\frac{1}{Z^\prime}\int\mathcal D[\bar \psi, \psi]e^{-S^\prime}d_\sigma(\tau)\bar d_\sigma(\tau^\prime)\label{eq_Green_function_d},\\
G_{d\bm k}^\sigma(\tau, \tau^\prime)=-\frac{1}{Z^\prime}\int\mathcal D[\bar \psi, \psi]e^{-S^\prime}d_\sigma(\tau)\bar c_{\bm k\sigma}(\tau^\prime)\label{eq_Green_function_dk}.
\end{gather}
Since the Green functions are shown to have a time translation symmetry \cite{Yamamoto24prep}, we can introduce the Fourier transformation for the Green functions as
\begin{gather}
G_d^\sigma(\tau,\tau^\prime)=G_d^\sigma(\tau-\tau^\prime)=\frac{1}{\beta}\sum_{\omega_n} e^{-i\omega_n(\tau-\tau^\prime)}G_d^\sigma(i\omega_n),\\
G_d^\sigma(i\omega_n)=\int_0^\beta e^{i\omega_n\tau} G_d^\sigma(\tau)d\tau\label{eq_Fourier_green},
\end{gather}
where $\omega_n=(2n+1)\pi/\beta$ with $n\in\mathbb Z$ is the Matsubara frequency for fermions and similar equations hold for $G_{d\bm k}^\sigma(\tau, \tau^\prime)$. We also introduce the Fourier transformation for the Grassmann fields and the Kronecker delta as
\begin{gather}
d_\sigma(\tau)=\frac{1}{\sqrt \beta}\sum_{\omega_n}d_\sigma(\omega_n)e^{-i\omega_n \tau},\label{eq_Fourier_d}\\
\bar d_\sigma(\tau)=\frac{1}{\sqrt \beta}\sum_{\omega_n}\bar d_\sigma(\omega_n)e^{i\omega_n \tau}\label{eq_Fourier_dbar},\\
\delta_{\omega_n, 0}=\frac{1}{\beta}\int_0^\beta d\tau e^{i\omega_n\tau}\label{eq_Fourier_Kronecker},
\end{gather}
respectively. The Fourier transformation for $c_{\bm k\sigma}(\tau)$ and $\bar c_{\bm k\sigma}(\tau)$ is introduced in a similar way as Eqs.~\eqref{eq_Fourier_d} and \eqref{eq_Fourier_dbar}.

We first calculate the Fourier component of the Green function in Eq.~\eqref{eq_Green_function_d} in order to evaluate Eq.~\eqref{eq_Green_d}. By integrating out fermion degrees of freedom in the reservoir in Eq.~\eqref{eq_SMF}, the partition function is rewritten as
\begin{align}
Z^\prime=\tilde Z\int\mathcal D[\bar d, d]e^{-S_\mathrm{eff}^\prime},
\end{align}
where
\begin{align}
S_\mathrm{eff}^\prime=&\int_0^\beta d\tau\sum_\sigma \bar d_\sigma(\tau)(\partial_\tau+\tilde E_d +\tilde \lambda)d_\sigma(\tau)\notag\\
&+\int_0^\beta d\tau d\tau^\prime \sum_\sigma \bar d_\sigma(\tau)\Sigma_d^\sigma(\tau, \tau^\prime)d_\sigma(\tau^\prime)\label{eq_SMFeff1},
\end{align}
is the effective action,
\begin{align}
\Sigma_d^\sigma(\tau, \tau^\prime)=\sum_{\bm k}|V_{\bm k d}|^2\bar{b} b G_{\bm k}^\sigma(\tau,\tau^\prime),\label{eq_self_impurity}
\end{align}
is the self-energy of the impurity, and
\begin{gather}
G_{\bm k}^\sigma(\tau, \tau^\prime)=-(\partial_\tau+\epsilon_{\bm k})^{-1}\label{eq_green_electron},
\end{gather}
is the Green function of fermions in the reservoir. We note that, though the coefficient 
\begin{align}
\tilde Z=e^{-\int_0^\beta d\tau \tilde \lambda (b_0^2-1)}
\prod_{\bm k \sigma} \det(\partial_\tau+\epsilon_{\bm k}) 
\end{align}
does not affect the calculation of $G_d^\sigma(\tau,\tau^\prime)$, the factor $\prod_{\bm k \sigma} \det(\partial_\tau+\epsilon_{\bm k})$ plays an important role to obtain $G_{d\bm k}^\sigma(\tau, \tau^\prime)$ [see Eq.~\eqref{eq_Green_kd2} below]. Since substituting the Fourier transform \eqref{eq_Fourier_green} and \eqref{eq_Fourier_d} into Eq.~\eqref{eq_SMFeff1} yields
\begin{align}
S^\prime_\mathrm{eff}=\sum_{\omega_n, \sigma}\bar d_\sigma(\omega_n)(-i\omega_n+\tilde E_d +\tilde \lambda+\Sigma_d^\sigma(i\omega_n))d_\sigma(\omega_n),
\label{eq_action_d}
\end{align}
where we have used Eq.~\eqref{eq_Fourier_Kronecker}, we arrive at the Fourier representation of the Green function as
\begin{align}
G_d^\sigma(i\omega_n)=&-\frac{1}{Z_\mathrm{eff}}\int \mathcal D[\bar d, d]d_\sigma(\omega_n) \bar d_\sigma(\omega_n) e^{-S_\mathrm{eff}^\prime}\notag\displaybreak[2]\\
=&\frac{1}{i\omega_n-\tilde E_d-\tilde \lambda-\Sigma_d^\sigma(i\omega_n)}.
\label{eq_Green_d_omega}
\end{align}
Here, we have introduced $Z_\mathrm{eff}=\int\mathcal D[\bar d, d]e^{-S_\mathrm{eff}^\prime}$ and the Fourier transform of Eqs.~\eqref{eq_self_impurity} and \eqref{eq_green_electron} as
\begin{gather}
\Sigma_d^\sigma(i\omega_n)=\sum_{\bm k}|V_{\bm k d}|^2\bar {b} b G_{\bm k}^\sigma(i\omega_n)\label{eq_SelfEnergy_d}, \displaybreak[2]\\
G_{\bm k}^\sigma(i\omega_n)=\frac{1}{i\omega_n-\epsilon_{\bm k}}.
\end{gather}

\begin{figure}[t]
\includegraphics[width=8.5cm]{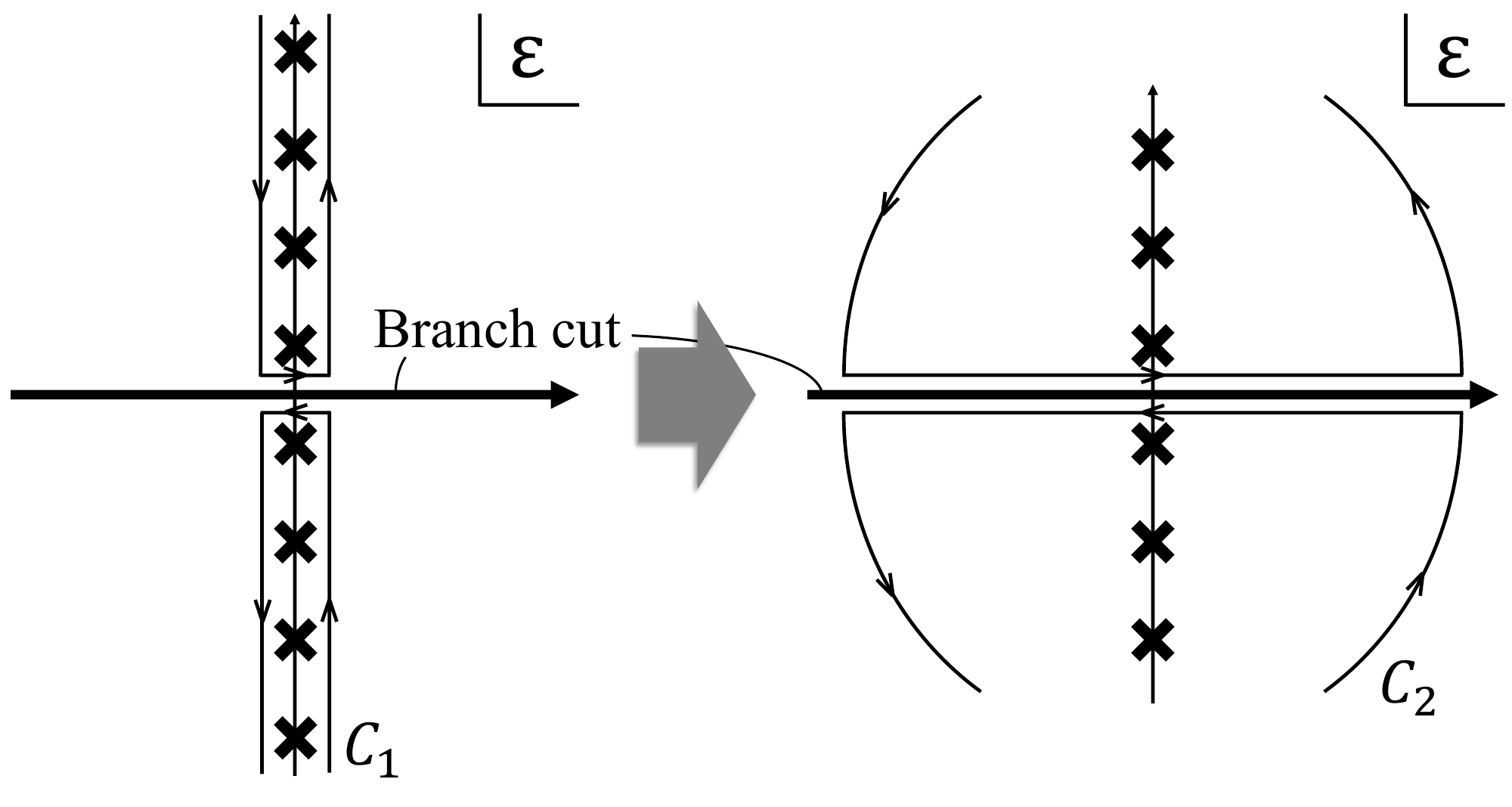}
\caption{\cred{Schematic figure of the contour integration with Green functions. There are no poles of $\tilde G_d^{R\sigma}(\epsilon)$ nor $\tilde G_d^{A\sigma}(\epsilon)$ inside the contour for $\Delta_b^\mathrm{Re}\pm\gamma^\prime>0$. Cross marks show the position of $i\omega_n$, where $\omega_n=(2n+1)\pi/\beta$ with $n\in\mathbb Z$ is the Matsubara frequency of fermions.}}
\label{fig_Contour}
\end{figure}

Now, we are ready to calculate Eq.~\eqref{eq_Green_d} by using $G_d^\sigma(i\omega_n)$ given in Eq.~\eqref{eq_Green_d_omega}. The analytic continuation of the impurity self-energy \eqref{eq_SelfEnergy_d} to the complex $\omega$ plane is calculated as 
\begin{align}
\Sigma_d^\sigma(\omega)&\sim V^2\bar {b}b \sum_{\bm k}\frac{1}{\omega-\epsilon_{\bm k}}\notag\displaybreak[2]\\
&=V^2\bar {b}b \int_{-\infty}^\infty d\epsilon \rho(\epsilon)\frac{1}{\omega-\epsilon}\notag\displaybreak[2]\\
&\sim V^2\bar {b}b \rho_0\int_{-\infty}^\infty d\epsilon \frac{\mathrm{Re}\omega-\epsilon-i\mathrm{Im}\omega}{(\mathrm{Re}\omega-\epsilon)^2+(\mathrm{Im}\omega)^2}\notag\displaybreak[2]\\
&=-i\bar {b}b \Delta \mathrm{sgn}(\mathrm{Im}\omega)
\label{eq_SelfEnergy2}
\end{align}
where the fermion reservoir is assumed to have a constant density of states $\rho_0=1/(2D)$ with the cutoff $D$, and we have introduced $\Delta=\pi \rho_0 V^2$ and assumed $|V_{\bm k d}|^2\sim V^2$. Then, we evaluate Eq.~\eqref{eq_Green_d} with the contour integration shown in Fig.~\ref{fig_Contour} by noting that there exists a branch cut along the real axis generated by the self-energy $\Sigma_d^\sigma(\omega)$ \cite{Coleman87}. Here, we assume that $\Delta_b^\mathrm{Re}\pm\gamma^\prime>0$ by introducing the real part of the renormalized hybridization $\Delta$ and (twice) the renormalized one-body loss rate as $\Delta_b^\mathrm{Re}\equiv\Delta\mathrm{Re}[\bar {b} b]$ and $\gamma^\prime\equiv\gamma/2-\mathrm{Im}\tilde \lambda$, respectively. The case of $\Delta_b^\mathrm{Re}\pm\gamma^\prime<0$ will be explained in the later section. The calculation proceeds as
\begin{align}
{}_L\langle d_\sigma^\dag d_\sigma\rangle_R
=&\frac{1}{\beta}\sum_{\omega_n} e^{i\omega_n0^+}G_d^\sigma(i\omega_n)\notag\\
=&-\frac{1}{2\pi i}\oint_{C_2(=C_1)}d\epsilon e^{\epsilon 0^+}\tilde G_d^\sigma(\epsilon)f(\epsilon)\notag\\
=&-\frac{1}{2\pi i}\int_{-\infty}^\infty d\epsilon f(\epsilon)\Bigg(\frac{1}{\epsilon-\tilde E_d-\tilde \lambda +i \bar{ b} b \Delta}\notag\displaybreak[2]\\
&-\frac{1}{\epsilon-\tilde E_d-\tilde \lambda -i \bar{b} b \Delta}\Bigg),
\label{eq_Green_d2}
\end{align}
where we have used the Fermi distribution function $f(\epsilon)=1/(e^{\beta\epsilon}+1)$ as it has poles at $i\omega_n$ and appropriately make the contour integration vanish at infinite distance on the arc in the contour $C_2$ shown in Fig.~\ref{fig_Contour}. Equation \eqref{eq_Green_d2} in the $\beta\to\infty$ limit is rewritten as
\begin{widetext}
\begin{align}
{}_L\langle d_\sigma^\dag d_\sigma\rangle_R
=&-\frac{1}{2\pi i}\int_{-\infty}^0 d\epsilon \left(\frac{\epsilon-E_d^\prime-\Delta_b^\mathrm{Im}-i(\Delta_b^\mathrm{Re}+\gamma^\prime)}{(\epsilon-E_d^\prime-\Delta_b^\mathrm{Im})^2 + (\Delta_b^\mathrm{Re}+\gamma^\prime)^2}-\frac{\epsilon-E_d^\prime+\Delta_b^\mathrm{Im}+i(\Delta_b^\mathrm{Re}-\gamma^\prime)}{(\epsilon-E_d^\prime+\Delta_b^\mathrm{Im})^2 + (\Delta_b^\mathrm{Re}-\gamma^\prime)^2}\right)\notag\displaybreak[2]\\
=&-\frac{1}{2\pi i}\Bigg[\frac{1}{2}\log\left((\epsilon-E_d^\prime-\Delta_b^\mathrm{Im})^2+(\Delta_b^\mathrm{Re}+\gamma^\prime)^2\right)
+i\tan^{-1}\left(\frac{E_d^\prime+\Delta_b^\mathrm{Im}-\epsilon}{\Delta_b^\mathrm{Re}+\gamma^\prime}\right)\notag\displaybreak[2]\\
&\quad-\frac{1}{2}\log\left((\epsilon-E_d^\prime+\Delta_b^\mathrm{Im})^2+(\Delta_b^\mathrm{Re}-\gamma^\prime)^2\right)
+i\tan^{-1}\left(\frac{E_d^\prime-\Delta_b^\mathrm{Im}-\epsilon}{\Delta_b^\mathrm{Re}-\gamma^\prime}\right)
\Bigg]_{-\infty}^0\notag\displaybreak[2]\\
=&-\frac{1}{4\pi i}\log\left(\frac{(E_d^\prime+\Delta_b^\mathrm{Im})^2+(\Delta_b^\mathrm{Re}+\gamma^\prime)^2}{(E_d^\prime-\Delta_b^\mathrm{Im})^2+(\Delta_b^\mathrm{Re}-\gamma^\prime)^2}\right)+\frac{1}{2}-\frac{1}{2\pi}\left[\tan^{-1}\left(\frac{E_d^\prime+\Delta_b^\mathrm{Im}}{\Delta_b^\mathrm{Re}+\gamma^\prime}\right)+\tan^{-1}\left(\frac{E_d^\prime-\Delta_b^\mathrm{Im}}{\Delta_b^\mathrm{Re}-\gamma^\prime}\right)\right],
\label{eq_Green_d3}
\end{align}
\end{widetext}
where we have defined the imaginary part of the renormalized hybridization $\Delta$ and the renormalized impurity level as $\Delta_b^\mathrm{Im}\equiv\Delta\mathrm{Im}(\bar {b} b)$ and $E_d^\prime\equiv E_d+\mathrm{Re}\tilde\lambda$, respectively. In Eq.~\eqref{eq_Green_d3}, we have taken the principal value of $\tan^{-1}x$ as $-\frac{\pi}{2}<\tan^{-1}x<\frac{\pi}{2}$. We see in the last line in Eq.~\eqref{eq_Green_d3} that the first term gives a nontrivial contribution to the imaginary part of ${}_L\langle d_\sigma^\dag d_\sigma\rangle_R$, which is a unique feature of the NH SB mean-field theory originating from the effect of dissipation. This fact is explicitly seen by rewriting Eq.~\eqref{eq_Green_d2} in the $\beta\to\infty$ limit as
\begin{align}
{}_L\langle d_\sigma^\dag d_\sigma\rangle_R=-\frac{1}{2\pi i}\int_{-\infty}^0d\epsilon(\tilde G_d^{R\sigma}(\epsilon)-\tilde G_d^{A\sigma}(\epsilon)) \in \mathbb{C},
\label{eq_numberLR}
\end{align}
where the retarded (advanced) Green function is defined by the analytic continuation of the Matsubara Green function as
\begin{align}
\tilde G_d^{R(A)\sigma}(\omega)=G_d^\sigma(i\omega_n\to\omega\pm i\eta)=\frac{1}{\omega-\tilde E_d-\tilde \lambda \pm i\bar{b}b\Delta}.
\end{align}
Because the parameters $\tilde E_d$, $\tilde \lambda$, $b$, and $\bar b$ are complex-valued, $\tilde G_d^{R\sigma}(\omega)$ and $\tilde G_d^{A\sigma}(\omega)$ are not complex conjugate to each other on the real-$\omega$ axis, rendering Eq.~\eqref{eq_Green_d2} to become a complex number. We note that, in the Hermitian limit, Eq.~\eqref{eq_numberLR} reduces to the well-known form given by the integration of the Lorentzian function as 
\begin{align}
\langle d_\sigma^\dag d_\sigma\rangle=-\frac{1}{\pi}\int_{-\infty}^0 d\epsilon \mathrm{Im}G_d^{R\sigma}(\epsilon)\in \mathbb R.
\end{align}

We see that Eq.~\eqref{eq_Green_d3} converges without introducing a cutoff $D$ in the integration. However, as shown in the calculation of Eq.~\eqref{eq_Green_kd} below, we must introduce a cutoff in order to guarantee the convergence of the SCEs \eqref{eq_saddle_b2} and \eqref{eq_saddle_lambda2}. Thus, we here rewrite Eq.~\eqref{eq_Green_d3} by introducing a cutoff $D$ as
\begin{widetext}
\begin{align}
{}_L\langle d_\sigma^\dag d_\sigma\rangle_R
=&-\frac{1}{2\pi i}\Bigg[\frac{1}{2}\log\left((\epsilon-E_d^\prime-\Delta_b^\mathrm{Im})^2+(\Delta_b^\mathrm{Re}+\gamma^\prime)^2\right)
+i\tan^{-1}\left(\frac{E_d^\prime+\Delta_b^\mathrm{Im}-\epsilon}{\Delta_b^\mathrm{Re}+\gamma^\prime}\right)\notag\displaybreak[2]\\
&\quad-\frac{1}{2}\log\left((\epsilon-E_d^\prime+\Delta_b^\mathrm{Im})^2+(\Delta_b^\mathrm{Re}-\gamma^\prime)^2\right)
+i\tan^{-1}\left(\frac{E_d^\prime-\Delta_b^\mathrm{Im}-\epsilon}{\Delta_b^\mathrm{Re}-\gamma^\prime}\right)
\Bigg]_{-D}^0\notag\displaybreak[2]\\
=&-\frac{1}{4\pi i}\log\left(\frac{(E_d^\prime+\Delta_b^\mathrm{Im})^2+(\Delta_b^\mathrm{Re}+\gamma^\prime)^2}{(D+E_d^\prime+\Delta_b^\mathrm{Im})^2+(\Delta_b^\mathrm{Re}+\gamma^\prime)^2}\right)
+\frac{1}{4\pi i}\log\left(\frac{(E_d^\prime-\Delta_b^\mathrm{Im})^2+(\Delta_b^\mathrm{Re}-\gamma^\prime)^2}{(D+E_d^\prime-\Delta_b^\mathrm{Im})^2+(\Delta_b^\mathrm{Re}-\gamma^\prime)^2}\right)\notag\displaybreak[2]\\
&-\frac{1}{2\pi}\left[\tan^{-1}\left(\frac{E_d^\prime+\Delta_b^\mathrm{Im}}{\Delta_b^\mathrm{Re}+\gamma^\prime}\right)+\tan^{-1}\left(\frac{E_d^\prime-\Delta_b^\mathrm{Im}}{\Delta_b^\mathrm{Re}-\gamma^\prime}\right)\right]\notag\displaybreak[2]\\
&+\frac{1}{2\pi}\left[\tan^{-1}\left(\frac{D+E_d^\prime+\Delta_b^\mathrm{Im}}{\Delta_b^\mathrm{Re}+\gamma^\prime}\right)+\tan^{-1}\left(\frac{D+E_d^\prime-\Delta_b^\mathrm{Im}}{\Delta_b^\mathrm{Re}-\gamma^\prime}\right)\right].
\label{eq_Green_d_cutoff}
\end{align}
\end{widetext}

\begin{figure}[t]
\includegraphics[width=8.5cm]{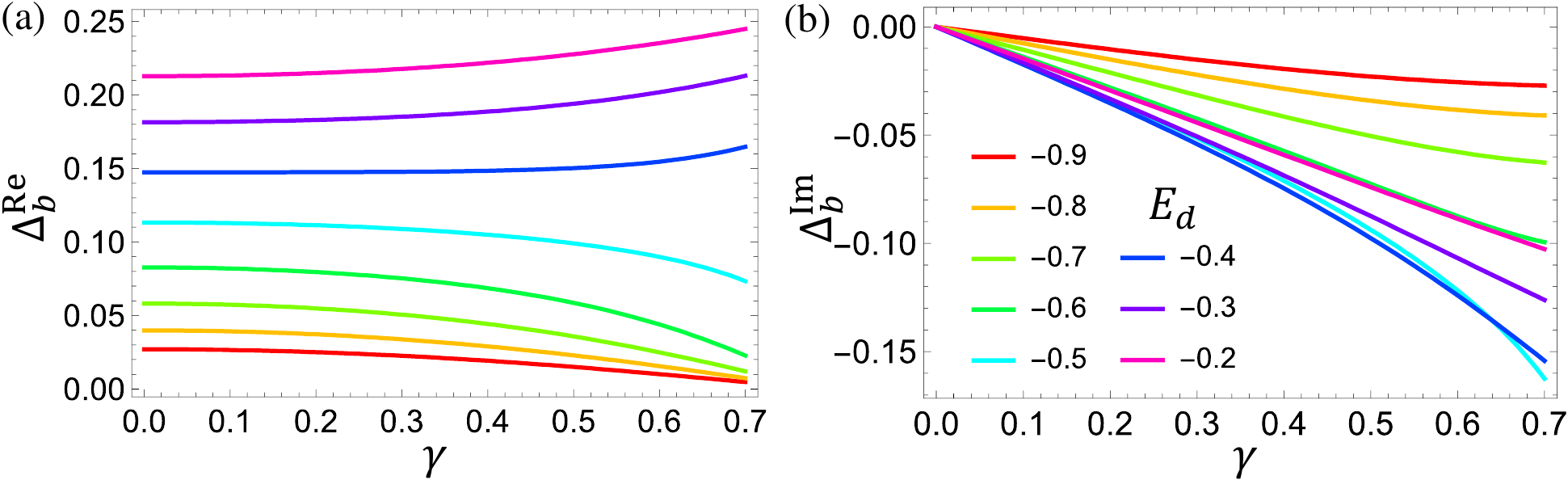}
\caption{\cred{Numerical results of (a) $\Delta_b^\mathrm{Re}$ and (b) $\Delta_b^\mathrm{Im}$ obtained from the SCEs \eqref{eq_self3} and \eqref{eq_self4}. The parameters are set to the same values as in Fig.~\ref{fig_KondoResonance}.}}
\label{fig_KondoResonance2}
\end{figure}

Next, we calculate $\sum_{\bm k}V_{\bm k d}{}_L\langle c_{\bm k \sigma}^\dag d_\sigma\rangle_R$ through Eq.~\eqref{eq_Green_kd} by evaluating the Green function \eqref{eq_Green_function_dk}. The Green function $G_{d\bm k}^\sigma(i\omega_n)$ is obtained by performing the path integral over the Grassmann variables as
\begin{align}
G_{d \bm k}^\sigma(i\omega_n)=&\frac{1}{Z^\prime}\int \mathcal D[\bar \psi, \psi]\bar c_{\bm k\sigma}(\omega_n)d_\sigma(\omega_n)e^{-S^\prime}\notag\displaybreak[2]\\
=&\frac{\int \mathcal D [\bar \psi_\sigma, \psi_\sigma] \bar c_{\bm k \sigma}(\omega_n)d_\sigma(\omega_n)e^{-S^\prime_\sigma}}{\int \mathcal D [\bar d_\sigma, d_\sigma]e^{-S_{\mathrm{eff}\sigma}^\prime}\prod_{\bm k, \omega_n}(-i\omega_n+\epsilon_{\bm k})}\notag\displaybreak[2]\\
=&\frac{\int \mathcal D [\bar c_{\bm k \sigma}, c_{\bm k \sigma}, \bar d_\sigma, d_\sigma] \bar c_{\bm k \sigma}(\omega_n)d_\sigma(\omega_n)e^{-S^{\prime\prime}_\sigma}}{\int \mathcal D [\bar d_\sigma, d_\sigma]e^{-S_{\mathrm{eff}\sigma}^{\prime\prime}}(-i\omega_n+\epsilon_{\bm k})}\notag\displaybreak[2]\\
=&\frac{V_{d\bm k} b}{(-i\omega_n+\tilde E_d+\tilde \lambda+\Sigma_d^\sigma(i\omega_n))(-i\omega_n+\epsilon_{\bm k})},
\label{eq_Green_kd2}
\end{align}
where
\begin{align}
S^\prime_\sigma=&\sum_{\omega_n}\Big\{\sum_{\bm k} \bar c_{\bm k \sigma}(\omega_n)(-i\omega_n+\epsilon_{\bm k})c_{\bm k \sigma}(\omega_n)\notag\displaybreak[2]\\
&+\bar d_\sigma(\omega_n)(-i\omega_n+\tilde E_d+\tilde \lambda)d_\sigma(\omega_n)\notag\displaybreak[2]\\
&+\sum_{\bm k}V_{\bm k d}\bar {b}\bar c_{\bm k \sigma}(\omega_n)d_\sigma(\omega_n)\notag\displaybreak[2]\\
&+\sum_{\bm k}V_{d\bm k}b \bar d_\sigma(\omega_n) c_{\bm k \sigma}(\omega_n)\Big\},\displaybreak[2]\\
S^\prime_{\mathrm{eff}\sigma}=&\sum_{\omega_n}\bar d_\sigma(\omega_n)(-i\omega_n+\tilde E_d +\tilde \lambda+\Sigma_d^\sigma(i\omega_n))d_\sigma(\omega_n),\displaybreak[2]\\
S_\sigma^{\prime\prime}=&\bar c_{\bm k \sigma}(\omega_n)(-i\omega_n+\epsilon_{\bm k})c_{\bm k \sigma}(\omega_n)\notag\displaybreak[2]\\
&+\bar d_\sigma(\omega_n)(-i\omega_n+\tilde E_d+\tilde \lambda+\Sigma_d^{\prime\sigma}(i\omega_n))d_\sigma(\omega_n)\notag\displaybreak[2]\\
&+V_{\bm k d}\bar {b}\bar c_{\bm k \sigma}(\omega_n)d_\sigma(\omega_n)+V_{d\bm k} b \bar d_\sigma(\omega_n) c_{\bm k \sigma}(\omega_n),\label{eq_S_sigma_primedouble}\displaybreak[2]\\
S^{\prime\prime}_{\mathrm{eff}\sigma}=&\bar d_\sigma(\omega_n)(-i\omega_n+\tilde E_d +\tilde \lambda+\Sigma_d^\sigma(i\omega_n))d_\sigma(\omega_n).
\end{align}
In the second line of Eq.~\eqref{eq_Green_kd2}, we have used the fact that the constant term and the $\sigma^\prime(\neq\sigma)$ component cancel out in forming the ratio of the numerator and the denominator, in the latter of which we have integrated out the fermion degrees of freedom in the reservoir. Then, in the third line in Eq.~\eqref{eq_Green_kd2}, the terms satisfying $\omega_n^\prime\neq\omega_n$ cancel out, and we have integrated out fermion degrees of freedom in the reservoir satisfying ${\bm k}^\prime\neq \bm k$ in the numerator. In Eq.~\eqref{eq_S_sigma_primedouble}, $\Sigma_d^{\prime\sigma}(i\omega_n)$ denotes the self-energy where the $\bm k^\prime$-sum in Eq.~\eqref{eq_SelfEnergy_d} is taken over ${\bm k}^\prime\neq \bm k$. By using the final line of Eq.~\eqref{eq_Green_kd2}, we obtain
\begin{align}
&\sum_{\bm k}V_{\bm k d}{}_L\langle c_{\bm k \sigma}^\dag d_\sigma\rangle_R\notag\displaybreak[2]\\
&=\frac{1}{\beta}\sum_{\omega_n}e^{i\omega_n0^+}G_{d}^\sigma(i\omega_n)\Sigma_d^\sigma(i\omega_n)\frac{1}{\bar {b}}\notag\displaybreak[2]\\
&=-\frac{1}{2\pi i}\oint_{C_2}d\epsilon e^{\epsilon 0^+}\tilde G_d^\sigma(\epsilon)\Sigma_d^\sigma(\epsilon) f(\epsilon)\frac{1}{\bar {b}}\notag\displaybreak[2]\\
&=-\frac{1}{2\pi i}\int_{-\infty}^\infty d\epsilon f(\epsilon)\Bigg(\frac{-ib \Delta}{\epsilon-\tilde E_d-\tilde \lambda +i \bar{b}b \Delta}\notag\displaybreak[2]\\
&\quad-\frac{ib \Delta}{\epsilon-\tilde E_d-\tilde \lambda -i \bar{b}b \Delta}\Bigg),
\label{eq_Green_kd_3}
\end{align}
where we have assumed $\Delta_b^\mathrm{Re}\pm\gamma^\prime>0$ and performed the contour integration shown in Fig.~\ref{fig_Contour}. Equation \eqref{eq_Green_kd_3} is evaluated in the $\beta\to\infty$ limit as
\begin{widetext}
\begin{align}
\sum_{\bm k}V_{\bm k d}{}_L\langle c_{\bm k \sigma}^\dag d_\sigma\rangle_R
=&\frac{b \Delta}{2\pi}\int_{-\infty}^0 d\epsilon \left(\frac{1}{\epsilon-\tilde E_d-\tilde \lambda +i \bar{b} b \Delta}+\frac{1}{\epsilon-\tilde E_d-\tilde \lambda -i \bar{b}b \Delta}\right)\notag\displaybreak[2]\\
=&\frac{b \Delta}{2\pi}\Bigg[\frac{1}{2}\log\left((\epsilon-E_d^\prime-\Delta_b^\mathrm{Im})^2+(\Delta_b^\mathrm{Re}+\gamma^\prime)^2\right)
+i\tan^{-1}\left(\frac{E_d^\prime+\Delta_b^\mathrm{Im}-\epsilon}{\Delta_b^\mathrm{Re}+\gamma^\prime}\right)\notag\displaybreak[2]\\
&\quad+\frac{1}{2}\log\left((\epsilon-E_d^\prime+\Delta_b^\mathrm{Im})^2+(\Delta_b^\mathrm{Re}-\gamma^\prime)^2\right)
-i\tan^{-1}\left(\frac{E_d^\prime-\Delta_b^\mathrm{Im}-\epsilon}{\Delta_b^\mathrm{Re}-\gamma^\prime}\right)
\Bigg]_{-\infty}^0.
\label{eq_Green_kd_4}
\end{align}
We emphasize that in Eq.~\eqref{eq_Green_kd_4}, the terms including $\tan^{-1}$ give a nontrivial imaginary contribution that does not exist in the Hermitian limit. As the terms containing logarithms do not converge, we have to introduce a cutoff $D$ in Eq.~\eqref{eq_Green_kd_4} \cite{Coleman87}. Therefore, we arrive at
\begin{align}
\sum_{\bm k}V_{\bm k d}{}_L\langle c_{\bm k \sigma}^\dag d_\sigma\rangle_R
=&\frac{b \Delta}{2\pi}\Bigg[\frac{1}{2}\log\left((\epsilon-E_d^\prime-\Delta_b^\mathrm{Im})^2+(\Delta_b^\mathrm{Re}+\gamma^\prime)^2\right)
+i\tan^{-1}\left(\frac{E_d^\prime+\Delta_b^\mathrm{Im}-\epsilon}{\Delta_b^\mathrm{Re}+\gamma^\prime}\right)\notag\displaybreak[2]\\
&\quad+\frac{1}{2}\log\left((\epsilon-E_d^\prime+\Delta_b^\mathrm{Im})^2+(\Delta_b^\mathrm{Re}-\gamma^\prime)^2\right)
-i\tan^{-1}\left(\frac{E_d^\prime-\Delta_b^\mathrm{Im}-\epsilon}{\Delta_b^\mathrm{Re}-\gamma^\prime}\right)
\Bigg]_{-D}^0\notag\displaybreak[2]\\
=&\frac{b \Delta}{4\pi}\log\left(\frac{(E_d^\prime+\Delta_b^\mathrm{Im})^2+(\Delta_b^\mathrm{Re}+\gamma^\prime)^2}{(D+E_d^\prime+\Delta_b^\mathrm{Im})^2+(\Delta_b^\mathrm{Re}+\gamma^\prime)^2}\right)
+\frac{b \Delta}{4\pi}\log\left(\frac{(E_d^\prime-\Delta_b^\mathrm{Im})^2+(\Delta_b^\mathrm{Re}-\gamma^\prime)^2}{(D+E_d^\prime-\Delta_b^\mathrm{Im})^2+(\Delta_b^\mathrm{Re}-\gamma^\prime)^2}\right)\notag\displaybreak[2]\\
&+\frac{ib \Delta}{2\pi}\left[\tan^{-1}\left(\frac{E_d^\prime+\Delta_b^\mathrm{Im}}{\Delta_b^\mathrm{Re}+\gamma^\prime}\right)-\tan^{-1}\left(\frac{E_d^\prime-\Delta_b^\mathrm{Im}}{\Delta_b^\mathrm{Re}-\gamma^\prime}\right)\right]\notag\displaybreak[2]\\
&-\frac{ib \Delta}{2\pi}\left[\tan^{-1}\left(\frac{D+E_d^\prime+\Delta_b^\mathrm{Im}}{\Delta_b^\mathrm{Re}+\gamma^\prime}\right)-\tan^{-1}\left(\frac{D+E_d^\prime-\Delta_b^\mathrm{Im}}{\Delta_b^\mathrm{Re}-\gamma^\prime}\right)\right].
\label{eq_Green_kd_cutoff}
\end{align}

Finally, the SCEs \eqref{eq_saddle_b2} and \eqref{eq_saddle_lambda2} are rewritten by using Eqs.~\eqref{eq_Green_d_cutoff} and \eqref{eq_Green_kd_cutoff} as
\begin{align}
&\tilde \lambda +\frac{\Delta}{2\pi}\left[\log\left(\frac{(E_d^\prime+\Delta_b^\mathrm{Im})^2+(\Delta_b^\mathrm{Re}+\gamma^\prime)^2}{(D+E_d^\prime+\Delta_b^\mathrm{Im})^2+(\Delta_b^\mathrm{Re}+\gamma^\prime)^2}\right)+\log\left(\frac{(E_d^\prime-\Delta_b^\mathrm{Im})^2+(\Delta_b^\mathrm{Re}-\gamma^\prime)^2}{(D+E_d^\prime-\Delta_b^\mathrm{Im})^2+(\Delta_b^\mathrm{Re}-\gamma^\prime)^2}\right)\right]\notag\displaybreak[2]\\
&+\frac{i\Delta}{\pi}\left[\tan^{-1}\left(\frac{E_d^\prime+\Delta_b^\mathrm{Im}}{\Delta_b^\mathrm{Re}+\gamma^\prime}\right)
-\tan^{-1}\left(\frac{E_d^\prime-\Delta_b^\mathrm{Im}}{\Delta_b^\mathrm{Re}-\gamma^\prime}\right)-\tan^{-1}\left(\frac{D+E_d^\prime+\Delta_b^\mathrm{Im}}{\Delta_b^\mathrm{Re}+\gamma^\prime}\right)+\tan^{-1}\left(\frac{D+E_d^\prime-\Delta_b^\mathrm{Im}}{\Delta_b^\mathrm{Re}-\gamma^\prime}\right)\right]
=0,
\label{eq_self3}\displaybreak[2]\\
&-\frac{1}{2\pi i}\left[\log\left(\frac{(E_d^\prime+\Delta_b^\mathrm{Im})^2+(\Delta_b^\mathrm{Re}+\gamma^\prime)^2}{(D+E_d^\prime+\Delta_b^\mathrm{Im})^2+(\Delta_b^\mathrm{Re}+\gamma^\prime)^2}\right)
-\log\left(\frac{(E_d^\prime-\Delta_b^\mathrm{Im})^2+(\Delta_b^\mathrm{Re}-\gamma^\prime)^2}{(D+E_d^\prime-\Delta_b^\mathrm{Im})^2+(\Delta_b^\mathrm{Re}-\gamma^\prime)^2}\right)\right]\notag\displaybreak[2]\\
&-\frac{1}{\pi}\left[\tan^{-1}\left(\frac{E_d^\prime+\Delta_b^\mathrm{Im}}{\Delta_b^\mathrm{Re}+\gamma^\prime}\right)+\tan^{-1}\left(\frac{E_d^\prime-\Delta_b^\mathrm{Im}}{\Delta_b^\mathrm{Re}-\gamma^\prime}\right)-\tan^{-1}\left(\frac{D+E_d^\prime+\Delta_b^\mathrm{Im}}{\Delta_b^\mathrm{Re}+\gamma^\prime}\right)-\tan^{-1}\left(\frac{D+E_d^\prime-\Delta_b^\mathrm{Im}}{\Delta_b^\mathrm{Re}-\gamma^\prime}\right)\right]+b_0^2=1,
\label{eq_self4}
\end{align}
\end{widetext}
where we have used the fact that the sum over $\sigma$ just gives the coefficient two because the Green functions do not depend on the spin. We note that Eq.~\eqref{eq_self3} is obtained by dividing Eq.~\eqref{eq_saddle_b2} by $b$ and the trivial solution $b=\bar b=0$ always satisfies the SCEs \eqref{eq_self3} and \eqref{eq_self4}, which are identical to Eq.~\eqref{eq_self5} by the transformation \eqref{eq_self3}$-i\Delta\times$\eqref{eq_self4} and \eqref{eq_self3}$+i\Delta\times$\eqref{eq_self4}. We show in Fig.~\ref{fig_KondoResonance2} the numerical results of $\Delta_b^\mathrm{Re}$ and $\Delta_b^\mathrm{Im}$ obtained from Eqs.~\eqref{eq_self3} and \eqref{eq_self4} as supplemental figures of Fig.~\ref{fig_KondoResonance}. For the deep impurity level $E_d$, we see that the behavior of $\Delta_b^\mathrm{Re}$ and $\Delta_b^\mathrm{Im}$ are quite similar to that of $\Delta_b^\mathrm{Re}+\gamma^\prime$ and $E_d^\prime+\Delta_b^\mathrm{Im}$ in Figs.~\ref{fig_KondoResonance}(c) and (e), reflecting that $E_d^\prime$ and $\gamma^\prime$ are almost pinned to zero as a result of the renormalization effect induced by strong correlations. As the impurity level is raised, we find that the one-body dissipation starts to govern the physics and $\Delta_b^\mathrm{Re}$ and $\Delta_b^\mathrm{Im}$ are renormalized so that their amplitude would be enhanced. These results indicate that strong correlations qualitatively change the nature of dissipation in the NH Kondo regime, while correlation effects are gradually suppressed as valence fluctuations occur.

\section{Detailed derivation of the NH Kondo scale}
\label{sec_Kondoscale}
We present the detailed derivation of the NH Kondo scale, for which we analytically evaluate $\Delta_b$ for the deep impurity level $E_d$ by using the SCEs \eqref{eq_self3} and \eqref{eq_self4}. By assuming that the cutoff $D$ is much larger than the other energy scales in the equation, we can rewrite Eqs.~\eqref{eq_self3} and \eqref{eq_self4} as a set of four equations given by
\begin{widetext}
\begin{gather}
E_d^\prime+\Delta_b^\mathrm{Im}=(\Delta_b^\mathrm{Re}+\gamma^\prime)\tan\left(\frac{\pi(\Delta_b^\mathrm{Re}+\gamma^\prime-\frac{\gamma}{2})}{2\Delta}\right),
\label{eq_self_an1}\\
E_d^\prime-\Delta_b^\mathrm{Im}=(\Delta_b^\mathrm{Re}-\gamma^\prime)\tan\left(\frac{\pi(\Delta_b^\mathrm{Re}-\gamma^\prime+\frac{\gamma}{2})}{2\Delta}\right),
\label{eq_self_an2}\\
(E_d^\prime+\Delta_b^\mathrm{Im})^2+(\Delta_b^\mathrm{Re}+\gamma^\prime)^2=D^2 \exp\left(\frac{\pi (E_d-E_d^\prime-\Delta_b^\mathrm{Im})}{\Delta}\right),
\label{eq_self_an3}\\
(E_d^\prime-\Delta_b^\mathrm{Im})^2+(\Delta_b^\mathrm{Re}-\gamma^\prime)^2=D^2 \exp\left(\frac{\pi (E_d-E_d^\prime+\Delta_b^\mathrm{Im})}{\Delta}\right).
\label{eq_self_an4}
\end{gather}
\end{widetext}
If the one-body loss rate is sufficiently small, i.e., $\gamma\ll\Delta$, we have
\begin{align}
\gamma^\prime, |\Delta_b^\mathrm{Im}|\ll E_d^\prime\ll\Delta_b^\mathrm{Re},
\label{eq_HerKondo}
\end{align}
for \cred{$|E_d|\gg\Delta$}. By evaluating Eqs.~\eqref{eq_self_an1}-\eqref{eq_self_an4} in this regime, we obtain
\begin{align}
\Delta_b^\mathrm{Re}+i\Delta_b^\mathrm{Im}\sim D\exp\left(\frac{\pi E_d}{2\Delta}\right)\equiv T_K^\mathrm{Her},
\label{eq_TK}
\end{align}
where $T_K^\mathrm{Her}$ is the Kondo temperature in the Hermitian case \cite{Coleman15}. When the dissipation strength $\gamma$ becomes large and comparable to $\Delta$, the contribution of $\Delta_b^\mathrm{Im}$ becomes significant and the physics is characterized by a complex energy scale $\Delta_b$. Then, by assuming that $\Delta_b^\mathrm{Re}\pm\gamma^\prime\ll\Delta$, we find that Eqs.~\eqref{eq_self_an1}-\eqref{eq_self_an4} are rewritten as
\begin{align}
\Delta_b^\mathrm{Re}\pm\gamma^\prime&\sim D\cos\left(\frac{\pi\gamma}{4\Delta}\right)\exp\left(\frac{\pi E_d}{2\Delta}\right),\label{eq_self_an5}\\
E_d^\prime\pm\Delta_b^\mathrm{Im}&\sim\mp D\sin\left(\frac{\pi\gamma}{4\Delta}\right)\exp\left(\frac{\pi E_d}{2\Delta}\right),\label{eq_self_an6}
\end{align}
where we have used the fact that the impurity level $E_d$ has to be deep enough in order that the assumption $\Delta_b^\mathrm{Re}\pm\gamma^\prime\ll\Delta$ should hold. From Eqs.~\eqref{eq_self_an5} and \eqref{eq_self_an6}, we obtain
\begin{align}
\gamma^\prime&\ll\Delta_b^\mathrm{Re},\label{eq_self_an7}\\
|E_d^\prime|&\ll|\Delta_b^\mathrm{Im}|,\label{eq_self_an8}
\end{align}
and arrive at the generalized NH Kondo scale given by
\begin{align}
\tilde T_K^\mathrm{NH}\equiv\Delta_b^\mathrm{Re}+i\Delta_b^\mathrm{Im}\sim D\exp\left(\frac{\pi\tilde E_d}{2\Delta}\right).
\label{eq_TKNHtilde_suppl}
\end{align}
As Eq.~\eqref{eq_TKNHtilde_suppl} is smoothly connected to $T_K^\mathrm{Her}$ given in Eq.~\eqref{eq_TK} in the $\gamma\to0$ limit, $\tilde T_K^\mathrm{NH}$ is regarded as a generalization of the Kondo temperature to the NH Kondo problem. Though Eq.~\eqref{eq_TKNHtilde_suppl} was obtained for relatively large dissipation, it gives a good approximation for the whole NH Kondo regime. Therefore, we find in the whole NH Kondo regime that the renormalized one-body loss rate $\gamma^\prime$ and the renormalized impurity level $|E_d^\prime|$ have a much smaller energy scale compared to $|\Delta_b|$ due to the renormalization effect. This is consistent with the numerical results obtained in Fig.~\ref{fig_KondoResonance}. Equation \eqref{eq_TKNHtilde_suppl} can be also interpreted from the generalized Kondo screening length $\tilde \xi$. We define the complex generalization of the Kondo screening length \cite{Affleck10, Borzenets20} as
\begin{align}
\tilde \xi =\frac{v_F}{\tilde T_K^\mathrm{NH}},
\end{align}
which can be regarded as the screening length of the complex Kondo cloud in real space. Here, $v_F$ is the Fermi velocity. Then, the phase transition at $\Delta^{\mathrm{Re}}_b \pm \gamma^\prime = 0$ indicates
\begin{align}
\frac{1}{\xi}\equiv\mathrm{Re}\frac{1}{\tilde\xi}=\frac{1}{v_F}\mathrm{Re}\tilde T_K^\mathrm{NH}=0.
\end{align}
Here, we have introduced the correlation length $\xi$ by separating $\exp(-L/\tilde \xi)$ into the decaying factor and the oscillating factor, the former of which is characterized by the correlation length as $\exp(-L/ \xi)$, where $L$ is the distance from the impurity. Thus, we find that the dissipation-induced phase transition is characterized by the divergence of the correlation length $\xi$.

\begin{figure}[t]
\includegraphics[width=8.5cm]{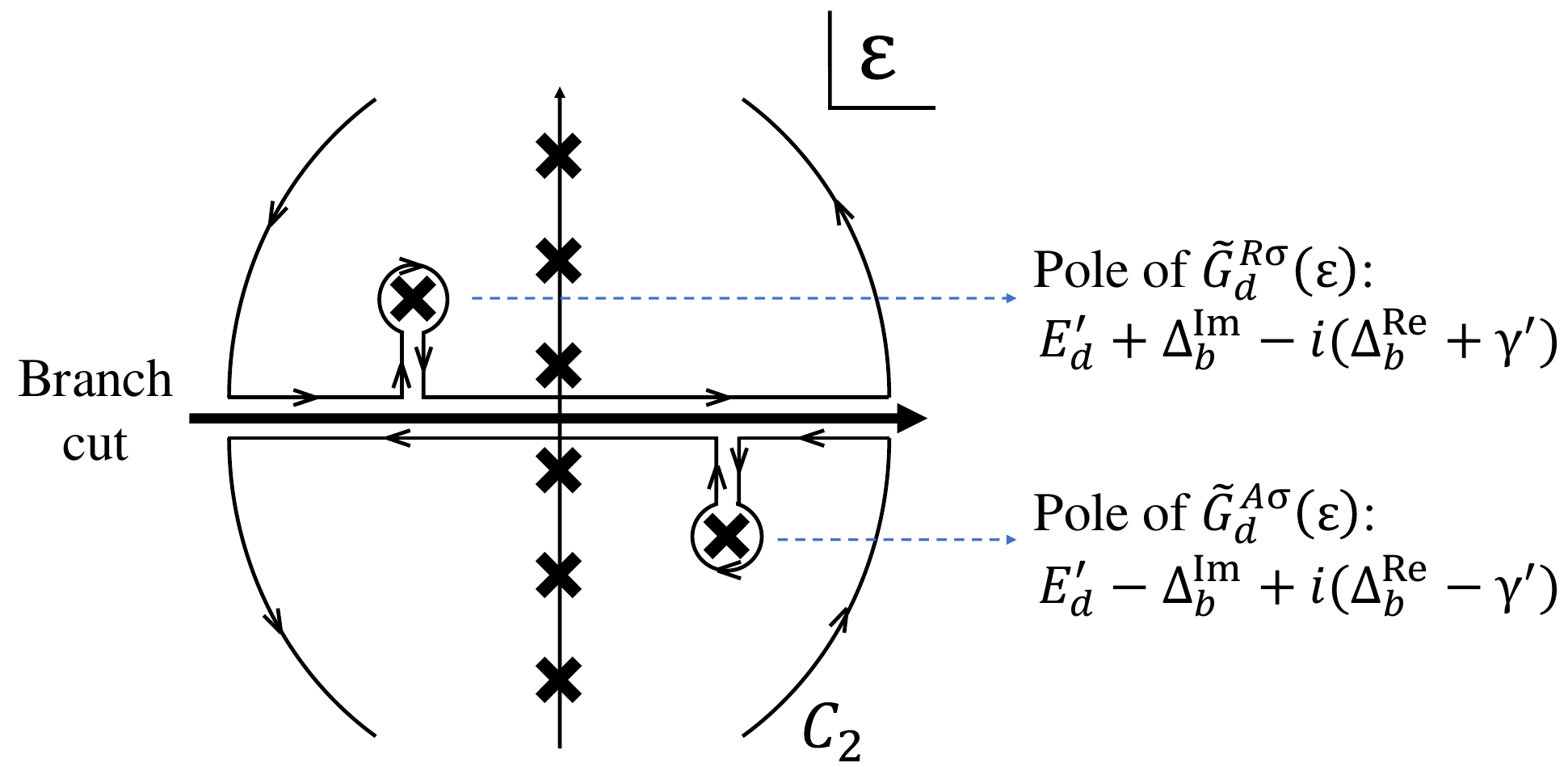}
\caption{\cred{Schematic figure of the contour integration with $\tilde G_d^{R(A)\sigma}(\epsilon)$. There exist poles of $\tilde G_d^{R\sigma}(\epsilon)$ and $\tilde G_d^{A\sigma}(\epsilon)$ inside the contour for $\Delta_b^\mathrm{Re}\pm\gamma^\prime<0$. Cross marks on the vertical axis show the position of $i\omega_n$, where $\omega_n=(2n+1)\pi/\beta$ with $n\in\mathbb Z$ is the Matsubara frequency of fermions.}}
\label{fig_ContourWithPole}
\end{figure}

\section{Modification of the self-consistent equation after the phase transition}
\label{sec_pole}
We explain how the SCE \eqref{eq_self5}, or Eqs.~\eqref{eq_self3} and \eqref{eq_self4}, is modified after the dissipation-induced phase transition. In the derivation of Eqs.~\eqref{eq_self3} and \eqref{eq_self4}, we have assumed that the poles of $\tilde G_d^{R(A)\sigma}(\epsilon)$ do not lie in the upper (lower) half complex plane by demanding 
\begin{align}
&\Delta_b^\mathrm{Re}+\gamma^\prime>0,\label{eq_nopole1}\\
&\Delta_b^\mathrm{Re}-\gamma^\prime>0.\label{eq_nopole2}
\end{align}
If either Eq.~\eqref{eq_nopole1} or \eqref{eq_nopole2} is not satisfied, the pole of $\tilde G_d^{R\sigma}(\epsilon)$ or $\tilde G_d^{A\sigma}(\epsilon)$ crosses the real axis and the contour integrations shown in Fig.~\ref{fig_Contour} should be modified. We note that $\Delta_b^\mathrm{Re}+\gamma^\prime$ and $\Delta_b^\mathrm{Re}-\gamma^\prime$ independently cross the real axis in general, and the phase transition occurs if either one of Eqs.~\eqref{eq_nopole1} and \eqref{eq_nopole2} breaks down. In what follows, we assume that the following relation holds after the phase transition:
\begin{align}
&\Delta_b^\mathrm{Re}+\gamma^\prime<0,\label{eq_pole1}\\
&\Delta_b^\mathrm{Re}-\gamma^\prime<0,\label{eq_pole2}
\end{align}
which will be justified below. For $\Delta_b^\mathrm{Re}\pm\gamma^\prime<0$, the sign of the Lorentzian function appearing in ${}_L\langle d_\sigma^\dag d_\sigma\rangle_R$ [see Eq.~\eqref{eq_Green_d3}] is reversed, and in this sense, the resonance width becomes negative. In fact, as the real part of ${}_L\langle d_\sigma^\dag d_\sigma\rangle_R$ given in Eq.~\eqref{eq_numberLR} is written as
\begin{widetext}
\begin{align}
\mathrm{Re}{}_L\langle d_\sigma^\dag d_\sigma\rangle_R
&=\int_{-\infty}^0 d\epsilon \left[\frac{1}{2\pi}\frac{\Delta_b^\mathrm{Re}+\gamma^\prime}{(\epsilon-E_d^\prime-\Delta_b^\mathrm{Im})^2 + (\Delta_b^\mathrm{Re}+\gamma^\prime)^2}+\frac{1}{2\pi}\frac{\Delta_b^\mathrm{Re}-\gamma^\prime}{(\epsilon-E_d^\prime+\Delta_b^\mathrm{Im})^2 + (\Delta_b^\mathrm{Re}-\gamma^\prime)^2}\right]\notag\\
&=-\frac{1}{2\pi}\int_{-\infty}^0 d\epsilon(\mathrm{Im}\tilde G_d^{R\sigma}(\epsilon)-\mathrm{Im}\tilde G_d^{A\sigma}(\epsilon)),
\end{align}
\end{widetext}
we find that the negative resonance width is regarded as the sign reversal of the Lorentzian function in the NH distribution $\mathrm{Re}{}_L\langle d_\sigma^\dag d_\sigma\rangle_R$ of impurity fermions. We note that such a negative resonance width gives the \credrev{negative value of the effective density of states}, which was also reported in the previous studies of noninteracting $\mathcal{PT}$-symmetric impurity systems \cite{Yoshimura20, Kulkarni22}. When Eqs.~\eqref{eq_pole1} and \eqref{eq_pole2} are satisfied, $\tilde G_d^{R(A)\sigma}(\epsilon)$ is merely defined from the analytic continuation of $G_d^\sigma(i\omega_n)$ to perform contour integrations, and their inverse Fourier transformation does not correspond to the ordinary time-ordered retarded (advanced) Green function \cite{Yamamoto24prep}. In this case, contour integrations should be performed along the modified path shown in Fig.~\ref{fig_ContourWithPole}. We note that such a contribution of the pole to the SCEs is reminiscent of that for the gap equation in NH BCS superfluidity \cite{Yamamoto19, Takemori24, Takemori24b}. Then, Eq.~\eqref{eq_Green_d2} is modified as
\begin{align}
{}_L\langle d_\sigma^\dag d_\sigma\rangle_R
=&\frac{1}{\beta}\sum_{\omega_n} e^{i\omega_n0^+}G_d^\sigma(i\omega_n)\notag\displaybreak[2]\\
=&-\frac{1}{2\pi i}\oint_{C_2}d\epsilon e^{\epsilon 0^+}\tilde G_d^\sigma(\epsilon)f(\epsilon)\notag\displaybreak[2]\\
=&-\frac{1}{2\pi i}\int_{-\infty}^\infty d\epsilon f(\epsilon)\Bigg(\frac{1}{\epsilon-\tilde E_d-\tilde \lambda +i \bar{b}b \Delta}\notag\\
&-\frac{1}{\epsilon-\tilde E_d-\tilde \lambda -i \bar{b}b \Delta}\Bigg)\notag\displaybreak[2]\\
&+f(\tilde E_d+\tilde \lambda -i \bar{b}b \Delta)+f(\tilde E_d+\tilde \lambda +i \bar{b}b \Delta).
\label{eq_Green_d2_pole}
\end{align}
Accordingly, Eq.~\eqref{eq_Green_kd_3} is modified as follows:
\begin{align}
&\sum_{\bm k}V_{\bm k d}{}_L\langle c_{\bm k \sigma}^\dag d_\sigma\rangle_R\notag\displaybreak[2]\\
&=\frac{1}{\beta}\sum_{\omega_n}e^{i\omega_n0^+}G_{d}^\sigma(i\omega_n)\Sigma_d^\sigma(i\omega_n)\frac{1}{\bar {b}}\notag\displaybreak[2]\\
&=-\frac{1}{2\pi i}\oint_{C_2}d\epsilon e^{\epsilon 0^+}\tilde G_d^\sigma(\epsilon)\Sigma_d^\sigma(\epsilon) f(\epsilon)\frac{1}{\bar {b}}\notag\displaybreak[2]\\
&=-\frac{1}{2\pi i}\int_{-\infty}^\infty d\epsilon f(\epsilon)\Bigg(\frac{-i b \Delta}{\epsilon-\tilde E_d-\tilde \lambda +i \bar{b} b \Delta}\notag\displaybreak[2]\\
&\quad-\frac{i b \Delta}{\epsilon-\tilde E_d-\tilde \lambda -i \bar{b} b \Delta}\Bigg)\notag\displaybreak[2]\\
&\quad-i b \Delta \left(f(\tilde E_d+\tilde \lambda -i \bar{b}b \Delta)-f(\tilde E_d+\tilde \lambda +i \bar{b}b \Delta)\right).
\label{eq_Green_kd_3_pole}
\end{align}
Thus, the SCEs given in Eqs.~\eqref{eq_self3} and \eqref{eq_self4} are modified to incorporate the contribution of the last line of Eqs.~\eqref{eq_Green_d2_pole} and \eqref{eq_Green_kd_3_pole}. Importantly, such a modification of the SCEs is necessary to retain a trivial solution $b_0=0$, where the impurity spin is decoupled from fermions in the reservoir. By evaluating the SCEs in the $\beta\to\infty$ limit, we see that the solutions with
\begin{align}
\Delta_b^\mathrm{Re}+\gamma^\prime>0,\\
\Delta_b^\mathrm{Re}-\gamma^\prime<0,
\end{align}
or
\begin{align}
\Delta_b^\mathrm{Re}+\gamma^\prime<0,\\
\Delta_b^\mathrm{Re}-\gamma^\prime>0
\end{align}
 are forbidden. We find that the existence of the nontrivial solution after the phase transition is permitted only when 
\begin{align}
\Delta_b^\mathrm{Re}\pm\gamma^\prime<0, \label{eq_solution_pole1}\\
E_d^\prime+\Delta_b^\mathrm{Im}<0, \label{eq_solution_pole2}\\
E_d^\prime-\Delta_b^\mathrm{Im}>0, \label{eq_solution_pole3}
\end{align}
or 
\begin{align}
\Delta_b^\mathrm{Re}\pm\gamma^\prime<0, \label{eq_solution_pole4}\\
E_d^\prime+\Delta_b^\mathrm{Im}>0, \label{eq_solution_pole5}\\
E_d^\prime-\Delta_b^\mathrm{Im}<0, \label{eq_solution_pole6}
\end{align}
is satisfied. From the numerical calculation, the solution after the phase transition is estimated to be in the region where Eqs.~\eqref{eq_solution_pole1}-\eqref{eq_solution_pole3} are satisfied. In this case, with the use of Eqs.~\eqref{eq_Green_d2_pole} and \eqref{eq_Green_kd_3_pole}, the SCEs \eqref{eq_self3} and \eqref{eq_self4} are modified in the $\beta\to\infty$ limit as
\begin{widetext}
\begin{align}
&\tilde \lambda +\frac{\Delta}{\pi}\log\left(\frac{(E_d^\prime+\Delta_b^\mathrm{Im})^2+(\Delta_b^\mathrm{Re}+\gamma^\prime)^2}{(D+E_d^\prime+\Delta_b^\mathrm{Im})^2+(\Delta_b^\mathrm{Re}+\gamma^\prime)^2}\right)
+\frac{2i\Delta}{\pi}\left[\tan^{-1}\left(\frac{E_d^\prime+\Delta_b^\mathrm{Im}}{\Delta_b^\mathrm{Re}+\gamma^\prime}\right)
-\tan^{-1}\left(\frac{D+E_d^\prime+\Delta_b^\mathrm{Im}}{\Delta_b^\mathrm{Re}+\gamma^\prime}\right)\right]-i\Delta b_0^2=3i\Delta,
\label{eq_self5pole}\displaybreak[2]\\
&\tilde \lambda+\frac{\Delta}{\pi}\log\left(\frac{(E_d^\prime-\Delta_b^\mathrm{Im})^2+(\Delta_b^\mathrm{Re}-\gamma^\prime)^2}{(D+E_d^\prime-\Delta_b^\mathrm{Im})^2+(\Delta_b^\mathrm{Re}-\gamma^\prime)^2}\right)
-\frac{2i\Delta}{\pi}\left[\tan^{-1}\left(\frac{E_d^\prime-\Delta_b^\mathrm{Im}}{\Delta_b^\mathrm{Re}-\gamma^\prime}\right)-\tan^{-1}\left(\frac{D+E_d^\prime-\Delta_b^\mathrm{Im}}{\Delta_b^\mathrm{Re}-\gamma^\prime}\right)\right]+i\Delta b_0^2=i\Delta.
\label{eq_self6pole}
\end{align}
\end{widetext}
Here, we find that Eqs.~\eqref{eq_self_an1}-\eqref{eq_self_an4} are not altered even when we include the contribution of poles, and thus the modification given in Eqs.~\eqref{eq_self5pole} and \eqref{eq_self6pole} would vanish for the large energy cutoff $D$.

\begin{figure*}[t]
\includegraphics[width=17cm]{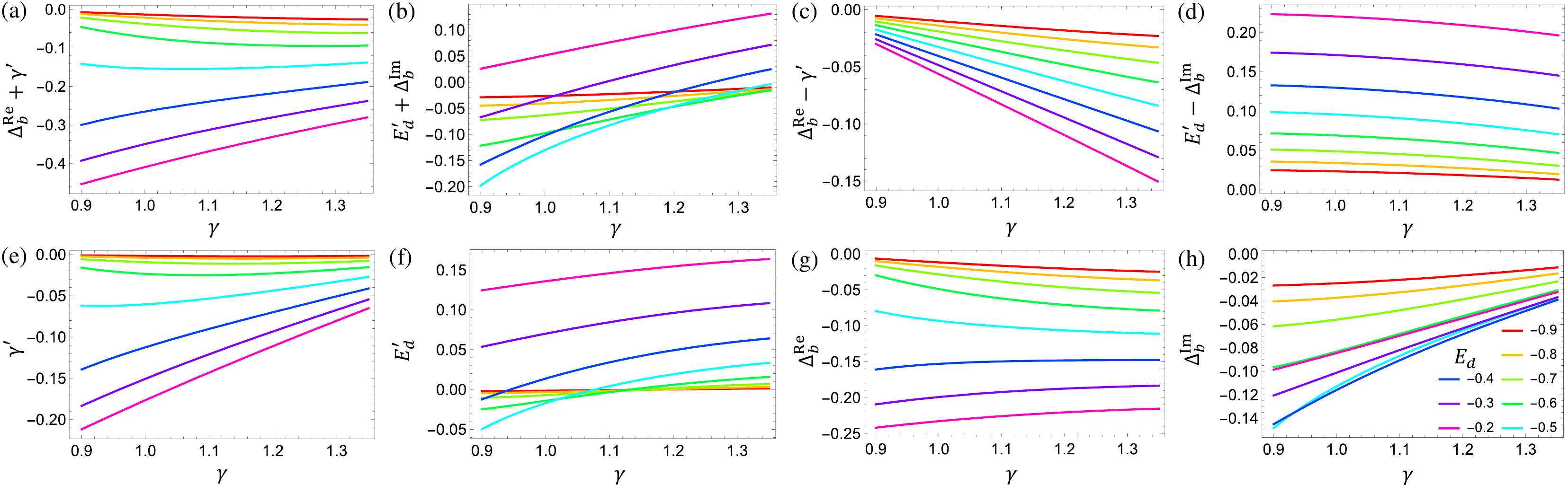}
\caption{\cred{Numerical solutions of the SCEs \eqref{eq_self5pole} and \eqref{eq_self6pole}. (a), (c) Renormalized resonance width and (b), (d) renormalized peak position read from $\tilde G_d^{R\sigma}(\omega)$ and $\tilde G_d^{A\sigma}(\omega)$. Results for different impurity levels are plotted by using different colors shown in the legend in (h). (e) The renormalized one-body loss rate and (f) the renormalized impurity level are almost pinned to zero for $|E_d|\gg\Delta$ even after the phase transition due to the renormalization effect. (g) $\Delta_b^\mathrm{Re}$ and (h) $\Delta_b^\mathrm{Im}$ are plotted for comparison. The parameters are set to the same values as those in Fig.~\ref{fig_KondoResonance}. We note that the results for $E_d^\prime+\Delta_b^\mathrm{Im}>0$ do not satisfy the condition \eqref{eq_solution_pole1}-\eqref{eq_solution_pole3} and should be omitted.}}
\label{fig_KondoResonanceWithPole}
\end{figure*}

Figure \ref{fig_KondoResonanceWithPole} shows the numerical solutions of the SCEs \eqref{eq_self5pole} and \eqref{eq_self6pole} after the phase transition, which is expected to occur at $\gamma\sim2\Delta$ from Eq.~\eqref{eq_phasetransition}. We note that, though we cannot obtain an analytical estimation of the transition point except for the NH Kondo regime, numerical results indicate that the entire regime ranging from the NH Kondo regime to the valence-fluctuation regime shows the phase transition almost at the same $\gamma$. Compared to Fig.~\ref{fig_KondoResonance} and Fig.~\ref{fig_KondoResonance2}, the behavior shown in Fig.~\ref{fig_KondoResonanceWithPole} seems to be reversed against the point (a) $(\Delta_b^\mathrm{Re}+\gamma^\prime,\gamma)=(0,2\Delta)$, (c) $(\Delta_b^\mathrm{Re}-\gamma^\prime,\gamma)=(0,2\Delta)$, (e) $(\gamma^\prime,\gamma)=(0,2\Delta)$, and (g) $(\Delta_b^\mathrm{Re},\gamma)=(0,2\Delta)$ and against the line $\gamma=2\Delta$ in (b), (d), (f), and (h). Though the region around $\gamma=2\Delta\sim 0.8$ is not shown due to numerical limitation, the behavior of the solution in this region can be inferred by comparing Fig.~\ref{fig_KondoResonanceWithPole} with Fig.~\ref{fig_KondoResonance} and Fig.~\ref{fig_KondoResonance2}. It seems that $\Delta_b^\mathrm{Re}+\gamma^\prime$ in Fig.~\ref{fig_KondoResonanceWithPole}(a), $\gamma^\prime$ in (e), and $\Delta_b^\mathrm{Re}$ in (g) jump from positive to negative at $\gamma=2\Delta\sim 0.8$ except for the Kondo limit, while $E_d^\prime+\Delta_b^\mathrm{Im}$ in (b), $\Delta_b^\mathrm{Re}-\gamma^\prime$ in (c), $E_d^\prime-\Delta_b^\mathrm{Im}$ in (d), $E_d^\prime$ in (f), and $\Delta_b^\mathrm{Im}$ in (h) seem to be smoothly connected at $\gamma=2\Delta$. When the phase transition occurs around $\gamma=2\Delta$, we see in Figs.~\ref{fig_KondoResonanceWithPole}(a) and (c) that the resonance width $\Delta_b^\mathrm{Re}\pm\gamma^\prime$ changes the sign from positive to negative, and its amplitude for the deep impurity level (e.g., see the plot for $E_d=-0.9$) is enhanced with increasing dissipation. On the other hand, in Figs.~\ref{fig_KondoResonanceWithPole}(b) and (d), we see that the renormalized peak position does not change the sign at the phase transition point $\gamma=2\Delta$ compared to Figs.~\ref{fig_KondoResonance}(e) and (f). Then, for example around $\gamma\sim0.9$ for $E_d=-0.9$ in Figs.~\ref{fig_KondoResonanceWithPole}(b) and (d), we see that the renormalized peak position $E_d^\prime+\Delta_b^\mathrm{Im}$ and $E_d^\prime-\Delta_b^\mathrm{Im}$ read off from $\tilde G_d^{R\sigma}(\epsilon)$ and $\tilde G_d^{A\sigma}(\epsilon)$ are just below and above the Fermi level, respectively. This means that the sign reversal of the NH Kondo peak obtained from $\Delta_b^\mathrm{Re}\pm\gamma^\prime$ does not affect its position in the energy level measured by $E_d^\prime\pm\Delta_b^\mathrm{Im}$. Moreover, in Figs.~\ref{fig_KondoResonanceWithPole}(e) and (f), we see for \cred{$|E_d|\gg\Delta$} that the renormalized one-body loss $\gamma^\prime$ and the renormalized impurity level $E_d^\prime$ are almost pinned to zero; they satisfy $|\Delta_b^\mathrm{Re}|\gg|\gamma^\prime|$ and $|\Delta_b^\mathrm{Im}|\gg |E_d^\prime|$. This demonstrates that, even after the phase transition, the metastable solution for the sufficiently deep impurity level still reflects the strong renormalization effects. We also find in Figs.~\ref{fig_KondoResonanceWithPole}(g) and (h) that both $\Delta_b^\mathrm{Re}$ and $\Delta_b^\mathrm{Im}$ take a negative value, which means that the real part of $b_0^2$ becomes negative. Then, from Eq.~\eqref{eq_saddle_lambda2}, we see that the negative sign of $\Delta_b^\mathrm{Re}$ let the real part of $\sum_\sigma{}_L\langle d_\sigma^\dag d_\sigma\rangle_R$ become greater than one \cite{Yoshimura20, Kulkarni22}. This is allowed because the ground state of the effective Hamiltonian $H_\mathrm{eff}$ does not conserve the quantities $d_\sigma^\dag d_\sigma$ and $b^\dag b$ though the total particle number is always conserved. Importantly, such exceeding of the expectation value $\mathrm{Re}\sum_\sigma{}_L\langle d_\sigma^\dag d_\sigma\rangle_R>1$ is prohibited in the Hermitian Kondo physics and demonstrates the intrinsic NH nature of the metastable solution. As shown in Fig.~\ref{fig_KondoTemperature}(b), the results plotted in Fig.~\ref{fig_KondoResonanceWithPole} are appropriately described by the analytical formula of the NH Kondo scale \eqref{eq_NHKondoTemp} for a sufficiently deep impurity level.

We next investigate what happens if we further increase the dissipation strength $\gamma$. So far, we have explained that the SCE should be modified to incorporate the contribution from poles of $\tilde G_d^{R(A)\sigma}(\omega)$ following Eqs.~\eqref{eq_solution_pole1}-\eqref{eq_solution_pole3} or Eqs.~\eqref{eq_solution_pole4}-\eqref{eq_solution_pole6}, but such a modification would vanish in the limit of the large energy cutoff $D$. Then, we can estimate the parameters for sufficiently deep impurity level $E_d$ by using Eqs.~\eqref{eq_self_an5}-\eqref{eq_self_an8}. It seems that there exists a regime that satisfies Eqs.~\eqref{eq_solution_pole4}-\eqref{eq_solution_pole6} for $4\Delta<\gamma<6\Delta$. However, we find that the numerical solution does not converge in this regime, and we need more sophisticated study. Also, since the resonance width $\Delta_b^\mathrm{Re}\pm\gamma^\prime$ and the peak position $E_d^\prime\pm\Delta_b^\mathrm{Im}$ are given by a periodic function with respect to $\gamma$ from Eqs.~\eqref{eq_self_an5} and \eqref{eq_self_an6}, the NH Kondo regime for $0<\gamma<2\Delta$ seems to reappear for $8\Delta<\gamma<10\Delta$. However, we cannot find any evidence of the reappearance from the present numerical result, which does not converge for $8\Delta<\gamma<10\Delta$. Thus, such reappearance of the NH Kondo phase for large $\gamma$ could possibly come from an artifact of the mean-field treatment of the SB theory. We conjecture that the NH Kondo phase does not reappear once the phase transition occurs at $\gamma=2\Delta$.

% Create the reference section using BibTeX:
\nocite{apsrev42Control}
\bibliographystyle{apsrev4-2}
\bibliography{NH_Anderson.bib}

\begin{thebibliography}{134}
\bibitem [{NHK()}]{NHKondo}%
  \BibitemOpen
  \href@noop {} {}\bibinfo {note} {\cred{Specifically, the renormalization
  group flow is analyzed in the NH Kondo model with two-body loss
  \cite{Nakagawa18}, in the Kondo limit of the $\mathcal{PT}$-symmetric
  Anderson model with NH hopping terms \cite{Lourenco18}, and in NH Luttinger
  liquids \cite{Han23}. Also, the NH quantum phase transitions are studied in
  the NH Kondo model with two-body loss or dephasing \cite{Nakagawa18,
  Kattel24, Hasegawa21}, and in $\mathcal{PT}$-symmetric spin chains with
  boundary dissipation \cite{Andrei24}. We emphasize that the impurity physics
  ranging from the Kondo effect to valence fluctuation has not been studied so
  far.}}\BibitemShut {Stop}%
\bibitem [{PTA()}]{PTAnderson}%
  \BibitemOpen
  \href@noop {} {}\bibinfo {note} {\cred{In Ref.~\cite{Kulkarni22}, a
  $\mathcal{PT}$-symmetric Anderson model with a NH Rashba spin-orbit coupling
  is studied and a quantum phase transition induced by the non-Hermiticity is
  found. Also, the relation between exceptional points and the phase transition
  is discussed mainly in the noninteracting limit $U=0$.}}\BibitemShut {Stop}%
\bibitem [{Foo()}]{Footnote}%
  \BibitemOpen
  \href@noop {} {}\bibinfo {note} {The Hilbert space is spanned by
  $\ket{\uparrow}=d_\uparrow^\dag|\Omega\rangle$,
  $\ket{\downarrow}=d_\downarrow^\dag|\Omega\rangle$, and
  $|0\rangle=b^\dag|\Omega\rangle$, where $|\Omega\rangle$ is a vacuum
  state.}\BibitemShut {Stop}%
\bibitem [{Mem()}]{Memo}%
  \BibitemOpen
  \href@noop {} {}\bibinfo {note} {The saddle-point equation for $b$, which is
  given by ${}_L\langle \frac{\delta S}{\delta
  b}\rangle_R=(-\partial_\tau+\tilde \lambda)\bar{b} +\sum_{\bm k
  \sigma}V_{d\bm k}{}_L\langle d_\sigma^\dag c_{\bm k \sigma}\rangle_R=0$, is
  equivalent to Eq.~\eqref{eq_saddle_b2}.}\BibitemShut {Stop}%
\bibitem [{Sym()}]{Symmetry}%
  \BibitemOpen
  \href@noop {} {}\bibinfo {note} {The effective Hamiltonian satisfies
  $H_\mathrm{eff}^\dag=H_\mathrm{eff}^*$ under the matrix representation in the
  Fock-state basis \cite{Yamamoto19, Takemori24, Takemori24b}, where we set
  $V_{\bm k d},V_{d\bm k}\in\mathbb R$ without loss of generality. Under the SB
  mean-field treatment, this symmetry constraint requires $b=\bar b\in \mathbb
  C$, and we arrive at Eqs.~\eqref{eq_b} and \eqref{eq_bbar} by multiplying the
  U(1) phase factor.}\BibitemShut {Stop}%
\bibitem [{pms()}]{pmsymbol}%
  \BibitemOpen
  \href@noop {} {}\bibinfo {note} {\cred{We note that double sign in equations
  should be read in the same order throughout the paper.}}\BibitemShut {Stop}%
\end{thebibliography}

\end{document}